\documentclass[12pt]{article}
\usepackage{epsfig}

\newcommand{\be}{\begin{equation}}
\newcommand{\ee}{\end{equation}}
\newcommand{\bea}{\begin{eqnarray}}
\newcommand{\eea}{\end{eqnarray}}
\newcommand{\beano}{\begin{eqnarrayno}}
\newcommand{\eeano}{\end{eqnarrayno}}
\newcommand{\nnm}{\nonumber}
\newcommand{\ds}{\displaystyle}

\newcommand{\tr}{\mathop{\rm tr}}

\newcommand{\ket}[1]{\mbox{$|#1\rangle$}}
\newcommand{\braket}[2]{\mbox{$\langle #1 | #2 \rangle$}}
\newcommand{\brakket}[3]{\mbox{$\langle #1 | #2 | #3 \rangle$}}

\newcommand{\D}{\mbox{
     \hbox to0pt{$D$\hss}\raisebox{1.7ex}{$\leftrightarrow$}}}
\newcommand{\Dright}{\mbox{
     \hbox to0pt{$D$\hss}\raisebox{1.7ex}{$\rightarrow$}}}
\newcommand{\Dleft}{\mbox{
     \hbox to0pt{$D$\hss}\raisebox{1.7ex}{$\leftarrow$}}}
\newcommand{\V}[1]{{\vec #1}}

\newcommand{\cj}{\dagger}
\newcommand{\note}[1]{}
\newcommand{\plus}{\mbox{$+$}}
\newcommand{\minus}{\mbox{$-$}}
\newcommand{\I}{{\rm i}}

\def\3{\ss}

\newcommand{\Dd}[1]{\mbox{
  \hbox to0pt{$D$\hss}\raisebox{1.7ex}{$\leftrightarrow$}$_{\!#1}$}}

\date{ }

\title{
\vbox{
\baselineskip12pt\normalsize
\hbox to\hsize{DESY 97-41\hfil}
\hbox to\hsize{HUP-EP-97/15\hfil}
\hbox to\hsize{UFTP 440/1997\hfil}
\hbox to\hsize{hep-lat/9703014\hfil}}
\vspace{1cm}
Pion and Rho Structure Functions from Lattice QCD}

\author{C. Best$^1$, M. G\"ockeler$^{1,2}$, R. Horsley$^{3}$, \\
        E.-M. Ilgenfritz$^3$,
        H. Perlt$^4$, P. Rakow$^5$, A. Sch\"afer$^1$, \\
        G. Schierholz$^{5,6}$, A. Schiller$^4$, S. Schramm$^7$
        \\[5mm]
         \small $^1$ Institut f\"ur Theoretische Physik,\\[-2mm]
             \small Johann Wolfgang Goethe--Universit\"at,\\[-2mm]
             \small D--60054 Frankfurt am Main, Germany \\[0mm]
        \small $^2$ Institut f\"ur Theoretische Physik,\\[-2mm]
             \small RWTH Aachen, D--52056 Aachen, Germany \\[0mm]
        \small $^3$ Institut f\"ur Physik, Humboldt-Universit\"at,\\[-2mm]
             \small D--10115 Berlin, Germany \\[0mm]
        \small $^4$ Institut f\"ur Theoretische Physik,\\[-2mm]
             \small Universit\"at Leipzig, D--04109 Leipzig, Germany\\[0mm]
        \small $^5$ Deutsches Elektronen-Synchrotron DESY,\\[-2mm]
             \small Institut f\"ur Hochenergiephysik and HLRZ,\\[-2mm]
             \small D--15735 Zeuthen, Germany \\[0mm]
        \small $^6$ Deutsches Elektronen-Synchrotron DESY,\\[-2mm]
             \small D--22603 Hamburg, Germany \\[0mm]
        \small $^7$ Gesellschaft f\"ur Schwerionenforschung GSI,\\[-2mm]
             \small D--64220 Darmstadt, Germany
        }

\begin{document}
\maketitle

\clearpage
\hspace{4cm}

\begin{abstract}
We calculate the lower moments of the deep-inelastic structure
functions of the pion and the rho meson on the lattice. Of particular 
interest to us are the spin-dependent structure functions of the rho. 
The calculations are done with Wilson 
fermions and for three values of the quark mass, so that we can perform
an extrapolation to the chiral limit.
\end{abstract}


\section{Introduction}

The deep-inelastic structure functions of nucleons and mesons are 
currently receiving a lot of attention, both experimentally and theoretically.
It has become possible to compute these structure functions 
from first principles within the framework of lattice 
QCD~\cite{Sachrajda,Nucleon}. This allows a 
quantitative test of QCD which goes beyond perturbation theory. 
The basis of the calculation is the operator product expansion which 
relates the moments of the structure functions to forward hadron 
matrix elements of certain local operators. Lattice simulations of these
matrix elements, combined with an appropriate calculation of the Wilson
coefficients, 
can in principle provide 
complete information of the quark and gluonic
structure of the hadronic states. The aim of this paper is 
to compute the structure functions of the pion and the rho meson. 
Following~\cite{Nucleon} we use Wilson fermions and work in the quenched
approximation, where internal quark loops are neglected.

The pion structure function, which so far was extracted from measurements 
of the Drell-Yan lepton-pair production cross section~\cite{Sutton}, is 
directly being measured at HERA at present~\cite{H1}, and we expect that 
new data will become available soon. Using the techniques described 
in~\cite{Craigie,Holtmann}, it should in principle be possible to measure 
the structure functions of the rho as well in the near future. 
Some of the spin-dependent structure functions, in particular, should be 
easy to separate from the dominant pion exchange process. 
 
But even without having any experimental data to compare with, 
the internal structure of the rho meson is an interesting 
subject to study on the lattice. Besides the structure functions already 
known from the nucleon, one finds new structure functions that
contain qualitatively new information which 
has no analog in the case of
spin-$\frac{1}{2}$ targets~\cite{Hoodbhoy}. We hope that this investigation
will lead to a better understanding of quark binding effects in hadrons. 

The structure functions of the rho are also of interest for the 
interpretation 
of photoproduction and two-photon inclusive cross sections, as the photon 
has a substantial hadronic component which to a good approximation can 
be described by the rho meson. 

For a spin-0 target like the pion, the kinematical framework is simpler 
than in the familiar nucleon case. The details for the case 
of a polarized spin-1 particle have been worked out by
Hoodbhoy et al.~\cite{Hoodbhoy}. The hadronic tensor, i.e.~the imaginary part 
of the forward current-hadron scattering amplitude,
\be \label{wmunu}
  W^{\mu\nu}(p,q,\lambda,\lambda')
  = \frac{1}{4\pi}
    \int {\rm d}^4x \, e^{\I q\cdot x} \,
    \brakket{p,\lambda'}{[j^\mu(x),j^\nu(0)]}{p,\lambda}
\ee
(with $\lambda$, $\lambda'$ labeling the polarization)
decomposes  into eight structure functions:
\bea \label{wmunu2}
  W_{\mu\nu} &=&
  - F_1 g_{\mu\nu}
  + F_2 \frac{p_\mu p_\nu}{\nu}
  - b_1 r_{\mu\nu}
  + \frac{1}{6} b_2 (s_{\mu\nu} + t_{\mu\nu} + u_{\mu\nu})
  + \frac{1}{2} b_3 (s_{\mu\nu} - u_{\mu\nu}) 
  \nnm\\ &&
  {}+ \frac{1}{2} b_4 (s_{\mu\nu} - t_{\mu\nu})
  + \I \frac{g_1}{\nu} \epsilon_{\mu\nu\lambda\sigma} q^\lambda s^\sigma
  \nnm\\ &&
  {}+ \I \frac{g_2}{\nu^2} \epsilon_{\mu\nu\lambda\sigma}
    q^\lambda ( p\cdot q s^\sigma - s \cdot q p^\sigma ) \quad,
\eea
where $\nu = p\cdot q$, and $r_{\mu\nu}$, $s_{\mu\nu}$, $t_{\mu\nu}$,
$u_{\mu\nu}$ are kinematical tensors \cite{Hoodbhoy} constructed from the
momentum transfer $q$ and the polarization vector $\epsilon$. The latter
 satisfies
$\epsilon \cdot p=0$, $\epsilon^2 = -m^2$, and
\be
  s^\sigma = \frac{-\I}{m^2} \epsilon^{\sigma\alpha\beta\tau}
             \epsilon_\alpha^* \epsilon_\beta p_\tau \quad,
\ee
with $m$ being the hadron mass. Here  $\epsilon^{\mu\nu\lambda\sigma}$
is the completely
antisymmetric tensor with $\epsilon^{0123}=-1$. 

The structure functions $F_1$, $F_2$, $g_1$, and $g_2$ play the same role 
as for a   spin-$\frac{1}{2}$  target. In the parton model, the structure of 
the hadron can be described by the probability $q^m_\uparrow(x)$ 
($q^m_\downarrow(x)$) of finding a quark with momentum fraction $x$ and 
spin up (down) along the direction of motion when the hadron is moving with 
infinite momentum and has spin projection $m=0,\pm1$. Symmetry implies
\begin{displaymath}
  q^1_\uparrow(x) = q^{-1}_\downarrow(x), \qquad
  q^1_\downarrow(x) = q^{-1}_\uparrow(x) \quad,
\end{displaymath}
\be
  q^0_\uparrow(x) = q^0_\downarrow(x) \quad,
\ee
so there remain only three independent parton   distribution functions 
$q^1_\uparrow(x)$, $q^1_\downarrow(x)$, and $q^0_\uparrow(x)$. 
In leading order, 
the single-flavor structure function 
$F^{(q)}_1(x)$ is one half of the probability to find a quark $q$ with
momentum fraction $x$, 
and $F^{(q)}_2(x)$ obeys the Callan-Gross relation:
\bea
  F^{(q)}_1(x) &=& \frac{1}{3} \left(
    q^1_\uparrow(x) + q^1_\downarrow(x) + q^0_\uparrow(x) \right)
           \quad,
  \nnm\\
  F^{(q)}_2(x) &=& 2x F^{(q)}_1(x) \quad.
\eea
In the complete structure function, $F^{(q)}_1(x)$ 
is weighted by the electric charge $Q_q$ of 
the quarks:
\bea
  F_1(x) = \sum_q Q_q^2 F^{(q)}_1(x) \quad.
\eea
In the following, we will only specify single-flavor structure functions and
omit the superscript $(q)$. 
For the mesons under consideration, the structure
functions are identical for both flavors.
The polarized structure
function $g_1(x)$ gives the fraction of spin carried by quarks:
\be
  g_1(x) = \frac{1}{2} \left[
           q^1_\uparrow(x) - q^1_\downarrow(x)
           \right ] \quad.
\ee
  The structure function $g_2(x)$ does not have a parton model interpretation.

The structure functions $b_1(x)$, $b_2(x)$, $b_3(x)$, and $b_4(x)$ are
particular to spin-1 targets as the kinematical   factors in eq.~(\ref{wmunu2})
involve the target polarization vector to second order, a feature that does
not occur for spin-$\frac{1}{2}$ targets. 
In parton model language $b_1(x)$ and $b_2(x)$ depend on the 
quark-spin averaged
distributions $q^m = \frac{1}{2}(q^m_\uparrow + q^m_\downarrow)$ only:
\bea \label{b12}
  b_1(x) &=& q^0(x) - q^1(x) \quad,\\
  b_2(x) &=& 2xb_1(x) \quad.
\eea
  Thus $b_1(x)$ and $b_2(x)$ measure the difference in parton distributions 
of an $m=1$ and 
$m=0$ target. This difference is due to the fact that in quantum field theory
any Lorentz boost changes the particle content of a state. These 
changes differ for different spin orientations relative to the boost 
direction. For a model discussion of $b_1$ see e.g.~\cite{Lech}. 

The paper is organized as follows. In Sec.\ \ref{secOpME}
we recall some
results from the operator product expansion concerning the relevant
operators and their matrix elements. Sec.\ \ref{sec3pt}
describes the method we
use to extract matrix elements from three-point functions. The
  lattice  implementation of this method is discussed in Sec.\ \ref{secEval}. 
Sec.\ \ref{secLatCont}
is devoted to questions of normalization and renormalization. In
Sec.\ \ref{secRes} we discuss our results, and Sec.\ \ref{secConc} presents our
conclusions. Appendix \ref{appConv} contains our conventions,
Appendices \ref{appOperators} and \ref{appSmearing}
describe some technicalities. The reader who is not interested in the 
computational details may skip Secs.\ \ref{sec3pt} and \ref{secEval}.

\section{Operators and moments of the structure functions}
\label{secOpME}
\label{secMoments}

The moments of structure functions can be related to the reduced matrix
elements of certain local operators between pion or
rho states. The local operators we consider are built from
$\gamma$ matrices and covariant derivatives and have the general form in
Minkowski space
\bea
  {\hat O}^{(M)\mu_1\cdots\mu_n}
  = \frac{1}{2^{n-1}} \,
    G_{ff'} \, \bar\psi_f \gamma^{\mu_1} \I\D^{\mu_2} \ldots
    \I\D^{\mu_n} \psi_{f'} \quad, \label{opdef} \\
  {\hat O}_5^{(M)\mu_1\cdots\mu_n}
  = \frac{1}{2^{n-1}} \,
    G_{ff'} \, \bar\psi_f \gamma^{\mu_1} \gamma_5 \I\D^{\mu_2} \ldots    
  \I\D^{\mu_n} \psi_{f'} \quad,
  \label{opdef5}
\eea
where $\psi$ is the quark field, and $G_{ff'}$ is a suitably chosen diagonal
flavor matrix. The
symmetrized derivative operators $\D$ are defined as
\be
  \D = \Dright - \Dleft \quad.
\ee

For a spin-0 particle, the momentum vector $p$ is 
the only quantity the matrix element can depend on, and the reduced matrix 
element $v_n$ is defined by
\be \label{opepion}
   \brakket{\V{p}}{\hat O^{(M)\{\mu_1 \cdots \mu_n\}}
                   - \mbox{traces}}{\V{p}}
   = 2v_n \, \left[ p^{\mu_1} \cdots p^{\mu_n} - \mbox{traces} \right] \quad.
\ee
The notation $\{\mu_1\cdots\mu_n\}$ denotes
symmetrization in the indices $\mu_1$, $\mu_2$, \dots, $\mu_n$. 
Expectation values of operators involving the $\gamma_5$ matrix vanish from 
symmetry considerations as the pion is a pseudoscalar particle.

For a spin-1 particle, the structure of the matrix elements is more
complicated due to the polarization degrees of freedom. 
Now both types of operators contribute\footnote{Note that we have corrected 
in (\ref{hoodbhoy2}) a misprint in \cite{Hoodbhoy}.}:
\begin{eqnarray} \label{hoodbhoy}
  \brakket{p,\lambda}{\hat O^{(M)\{\mu_1\ldots\mu_n\}}
                       - \mbox{traces}}{p,\lambda} &=&
  2S \Big[ a_n \, p^{\mu_1} \cdots p^{\mu_n} \nnm\\&& \hskip-35mm
          {}+d_n \, (\epsilon^{*\mu_1}(\V{p},\lambda)
                     \epsilon^{\mu_2}(\V{p},\lambda)
                   - \frac{1}{3} p^{\mu_1} p^{\mu_2})
               \, p^{\mu_3} \cdots p^{\mu_n} \Big] \quad, \\
  \brakket{p,\lambda}{\hat O_5^{(M)\{\mu_1\ldots\mu_n\}} 
                       - \mbox{traces}}{p,\lambda} && \nnm\\ && \hskip-35mm
  = \frac{2\I}{m^2} \, S \left[ r_n \, \epsilon^{\rho\sigma\tau\mu_1}
                  \epsilon^*_\rho(\V{p},\lambda) 
                  \epsilon_\sigma(\V{p},\lambda) 
                  p_\tau p^{\mu_2}
                  \cdots p^{\mu_n} \right] .
  \label{hoodbhoy2}
\end{eqnarray}
$S$ denotes symmetrization in the indices $\mu_1$, \dots, $\mu_n$ and
removal of traces. The reduced matrix elements are $a_n$ for the
polarization-averaged contribution, $d_n$ for the polarized contributions,
and $r_n$ for the operators involving $\gamma_5$.


By performing an operator product expansion of (\ref{wmunu}), reduced 
matrix elements of local operators can be related to moments of the 
structure functions. We define the $n$-th moment of a function $f(x)$ as
\note{Standardliteratur zitieren?}
\be
  M_n(f) = \int_0^1 x^{n-1} f(x) \, {\rm d}x \quad.
\ee
One then finds to leading order, which is twist two, the following
representation of the moments of the 
pion structure functions:
\bea \label{momentspion}
  2 M_n(F_1) &=& C^{(1)}_n \, v_n \quad, \nnm\\
  M_{n-1}(F_2) &=& C^{(2)}_n \, v_n \quad; 
\eea
for the rho structure functions one obtains \cite{Hoodbhoy}:
\bea \label{moments}
  2 M_n(F_1) &=& C^{(1)}_n \, a_n \quad, \nnm\\
  M_{n-1}(F_2) &=& C^{(2)}_n \, a_n \quad, \nnm\\
  2 M_n(b_1) &=& C^{(1)}_n \, d_n \quad, \nnm\\
  M_{n-1}(b_2) &=& C^{(2)}_n \, d_n \quad, \nnm\\
  2 M_n(g_1) &=& C^{(3)}_n \, r_n \quad,
\eea
where $C^{(k)}_n = 1 + O(\alpha_s)$ are the Wilson coefficients of the 
operator product expansion. These relations hold for even $n$, except for the 
last one, which is valid for odd $n$. However, since we are calculating in 
the quenched approximation, we are allowed to make use of these formulas
for all $n$ keeping in mind that our results can be meaningfully compared 
only with the non-singlet valence quark distribution. 

In the case of the 
pion, the moments of the quark distribution are given by
\be
  \langle x^{n-1} \rangle = v_n \quad,
\ee
while for the rho they are related to the matrix elements $a_n$:
\be
  \langle x^{n-1} \rangle = a_n \quad.
\ee

\section{Three-point functions and matrix elements}
\label{sec3pt}

In order to calculate the reduced matrix elements on the lattice, we
must calculate the expectation values of local operators of the form
(\ref{opdef}) and (\ref{opdef5}). To this end, we first need the
connection between the Minkowski operators and those in Euclidean
space. Defining Euclidean operators by
\begin{eqnarray}
 \label{eucopdef}
  {\hat O}^{(E)}_{\mu_1\cdots\mu_n}
  &=& G_{ff'} \, \frac{1}{2^{n-1}} \,
    \bar\psi_f \gamma^{(E)}_{\mu_1} \D^{(E)}_{\mu_2} \ldots
    \D^{(E)}_{\mu_n} \psi_{f'} \quad,\\
  {\hat O}^{(E)5}_{\mu_1\cdots\mu_n}
  &=& G_{ff'} \, \frac{1}{2^{n-1}} \,
    \bar\psi_f \gamma^{(E)}_{\mu_1} \gamma_5 \D^{(E)}_{\mu_2} \ldots
    \D^{(E)}_{\mu_n} \psi_{f'} \quad,
\end{eqnarray}
we obtain the following relation to the operators in Minkowski space:
\begin{equation}
 \label{euclo}
  \hat O^{(M)\mu_1\cdots \mu_n}
  = (-) \, (-)^{n_4+n_5} \, (-\I)^{n_{123}} \,
    \hat O^{(E)}_{\mu_1\cdots \mu_n} \quad,
\end{equation}
where $n_4$ is the number of time-like indices, $n_{123}$ the number of spatial
indices, and $n_5=1$ if the operator carries a $\gamma_5$ matrix. 
For our Euclidean conventions, see Appendix \ref{appConv}.

Lattice operators with the appropriate continuum behavior can be constructed
from the Wilson fermion fields by considering their symmetry properties
under the hypercubic group $H(4)$ \cite{Groupth}. The operators we have
chosen and their relation to the reduced matrix elements are listed in
Appendix \ref{appOperators}.

The required expectation values of our operators are extracted from
ratios of two- and three-point functions. The three-point functions we
consider are of the general form
\be 
  \langle \eta(t,\V{p}) \,
          {\cal O}(\tau) \, \eta^\cj(0,\V{p}) \rangle
\ee
where $\eta(t,\V{p})$ is the sink operator for a particle moving with 
momentum $\V{p}$ in time slice $t$, and $\eta^\cj(0,\V{p})$ is the 
corresponding source at time slice $t=0$. These operators are required to 
have the correct symmetry properties for the particles in question 
and their corresponding Hilbert space operators $\hat \eta(\V{p})$ should
create the desired particles from the vacuum with nonzero amplitude. 
${\cal O}(\tau)$ represents the operator $\hat O$ whose expectation value 
is to be calculated.

For the pion we write
\be
  \brakket{0}{\hat \eta(\pi;\V{p})}{\pi;\V{p}}
  = \sqrt{Z_\pi} \quad,
\ee
while for the rho there are three different particle states and
correspondingly three different operators arranged in a vector
$\eta_i(\rho;t,\V{p})$ that satisfy
\be \label{eta}
  \brakket{0}{\hat \eta_j(\rho;\V{p})}{\rho;\V{p},\lambda}
  = \sqrt{Z_\rho} \, \epsilon_j(\V{p},\lambda)
\ee
up to lattice artifacts (see Appendix \ref{appConv} for the definition of 
the polarization vectors $\epsilon_i$). The correlation function for the 
rho depends on the polarization vectors:
\be \label{corr1}
  C_{jk} = \langle \eta_j(\rho;t,\V{p}) \,
          {\cal O}(\tau) \, \eta_k^\cj(\rho;0,\V{p}) \rangle \quad.
\ee

In order to relate (\ref{corr1}) to the matrix elements we are interested
in, we express this correlation function
in terms of traces involving the transfer matrix $\hat S$:
\bea
\noalign{$\displaystyle
  \langle\eta_j(\rho;t,\V{p})\, {\cal O}(\tau)\, \eta^\cj_k(\rho;0,\V{p})
  \rangle = $}
  &=& \left\{ \begin{array}{cl}
    \tr \left[ \hat S^{T-t} \hat\eta_j(\rho;\V{p}) \hat S^{t-\tau} \hat O
               \hat S^\tau \hat\eta^\cj_k(\rho;\V{p}) \right] 
               & T \ge t \ge \tau \ge 0 \quad, \\[5pt]
    \tr \left[ \hat S^{T-\tau} \hat O \hat S^{\tau-t} \hat\eta_j(\rho;\V{p})
               \hat S^{t} \hat\eta^\cj_k(\rho;\V{p}) \right ] 
               & T \ge \tau \ge t \ge 0 \quad.
    \end{array} \right.
\label{tfm}\eea 
Here $T$ denotes the time extent of our lattice whose spacing is put equal
to $1$. Then we insert a complete set of orthonormal eigenstates of the
transfer matrix.
If the time differences are chosen sufficiently large, we can restrict 
ourselves to the lowest contributing states $\ket{\V{p},\lambda}$,
$\lambda$ labeling the three degenerate polarization states of the
rho. For the first case in (\ref{tfm}) one obtains
\bea \label{tfm2} C_{jk}^{(1)} &=&
  \sum_{\lambda \lambda'}
  \brakket{0}{\hat\eta_j(\rho;\V{p})}{\rho;\V{p},\lambda} \,
  \brakket{\rho;\V{p},\lambda}{\hat O}{\rho;\V{p},\lambda'} \,
  \brakket{\rho;\V{p},\lambda'}{\hat\eta_k(\rho;\V{p})}{0} \,
  e^{-Et} \nnm\\
  &=&
  Z_\rho \sum_{\lambda \lambda'}
  \epsilon_j(\V{p},\lambda) \, \epsilon^*_k(\V{p},\lambda') \,
  \brakket{\rho;\V{p},\lambda}{\hat O}{\rho;\V{p},\lambda'} \, e^{-Et} \nnm\\
  &=&
  Z_\rho \, m^2 \, T_{jk} \, e^{-Et} \quad,
\eea
\note{Need to say something about $C_2$.}
where $T_{jk}$ is the matrix element between Cartesian states,
\be
  T_{jk} = \frac{1}{m^2} \,\sum_{\lambda\lambda'} \epsilon_j(\V{p},\lambda) 
           \epsilon_k^*(\V{p},\lambda')
           \brakket{\rho;\V{p},\lambda}{\hat O}{\rho;\V{p},\lambda'} \quad.
\ee
In the second case, one finds an additional sign factor:
\be \label{c2sign}
  C_{jk}^{(2)} = (-1)^{n_4 + n_5} \, Z_\rho \, m^2 \, T_{jk} \, e^{-E(T-t)} 
  \quad.
\ee

To calculate all $C_{jk}$ components for a given momentum would be expensive
in computer time. Choosing the momentum in $1$-direction, we have restricted
ourselves to the components $C_{33}$ and $C_{32}$. If
$\brakket{\rho;\V{p},\plus}{\hat O}{\rho;\V{p},\minus}=0$ and
$\brakket{\rho;\V{p},\plus}{\hat O}{\rho;\V{p},\plus} =
\brakket{\rho;\V{p},\minus}{\hat O}{\rho;\V{p},\minus}$, then
\be \label{sign2}
  \brakket{\rho;\V{p},\plus}{\hat O}{\rho;\V{p},\plus} = T_{33} \quad, \qquad
  T_{32} = 0 \quad,
\ee
whereas for $\brakket{\rho;\V{p},\plus}{\hat
O}{\rho;\V{p},\plus} = - \brakket{\rho;\V{p},\minus}{\hat O}{\rho;\V{p},\minus}$,
\be \label{sign3}
  \brakket{\rho;\V{p},\plus}{\hat O}{\rho;\V{p},\plus} = -\I T_{32} \quad,\qquad
  T_{33} = 0 \quad.
\ee
The latter case is relevant to spin-dependent operators.   To satisfy
\[\brakket{\rho;\V{p},\plus}{\hat O}{\rho;\V{p},\minus} = 0\quad,\]  it is
sufficient that the operators commute with rotations in the plane transverse
to $\V{p}$. This has also motivated our choice of operators.

The factors that do not depend on the operator $\hat O$ can be eliminated 
by taking the ratio of (\ref{corr1}) to another correlator, e.g.~the 
two-point correlator
\bea \label{twopoint}
  C(t) &=& \sum_j \langle\eta_j(\rho;t,\V{p})\, 
           \eta^\cj_j(\rho;0,\V{p})\rangle\nnm\\
  &=& \sum_{\lambda} \Big[
  \brakket{0}{\hat\eta_j(\rho;\V{p})}{\rho;\V{p},\lambda} \,
  \brakket{\rho;\V{p},\lambda}{\hat\eta_k^\cj(\rho;\V{p})}{0} \,
  e^{-Et}
  \nnm \\ &&
  +
  \brakket{\rho;\V{p},\lambda}{\hat\eta_j(\rho;\V{p})}{0} \,
  \brakket{0}{\hat\eta_k^\cj(\rho;\V{p})}{\rho;\V{p},\lambda} \,
  e^{-E(T-t)} \Big] \quad.
\eea
Using the relations (\ref{eta}) and (\ref{epsilonsum}), this reduces to
\be
  C(t) =  Z_\rho \, (2 m^2 + E^2) \, \left( e^{-Et} + e^{-E(T-t)} \right)
  \quad.
\ee

We therefore arrive at the following relation between the ratio of a three-
to a two-point correlation function and the expectation value of the 
corresponding operator, valid for $t\gg\tau\gg 0$:
\bea 
  R_{jk} &=& \frac{\langle \eta_j(\rho;t,\V{p}) \,
                            {\cal O}(\tau) \,
                            \eta_k^\cj(\rho;0,\V{p}) \rangle
                     }{\sum_l \langle \eta_l(\rho;t,\V{p})
                                      \eta_l^\cj(\rho;0,\V{p}) \rangle} \nnm\\
  &=& 
    \frac{1}{2 + E^2/m^2} \, 
    \frac{e^{-Et}}{e^{-Et} + e^{-E(T-t)}}
      T_{jk} \quad. \label{ratio}
\eea
For $T \gg \tau \gg t \gg 0$, we get an analogous equation with the
additional sign factor from (\ref{c2sign})
and with $t$ replaced
by $T-t$. For $t=T/2$, 
which is the choice in our numerical work, (\ref{ratio}) gives
\be \label{ratio2}
  R_{jk} = 
    \frac{1}{2 + E^2/m^2} \, 
    \frac{1}{2} \,
      T_{jk} \quad.
\ee
The ratio may still depend on $\tau$ due to contributions from the higher
states neglected in (\ref{tfm2}). By searching for plateaus in the 
$\tau$-dependence, one can extract the value of the ratio with the smallest 
contamination from higher states.

In the case of the pion, there is no polarization, and the relation
(\ref{ratio2}) reduces to
\be \label{ratiopion}
  R = \frac{\brakket{\pi;\V{p}}{\hat O}{\pi;\V{p}}}{2} \quad.
\ee


\section{Evaluation of three-point functions on the lattice}
\label{secEval}

The actual form of the three-point correlator is given by
\be
  \langle \eta^F_\Gamma(t,\V{p}) \, {\cal O}^G(\tau) \,
          \eta^{F'}_{\Gamma'}(0,-\V{p}) \rangle \quad .
\ee
Here we explicitly indicate the flavor matrices $F$, $F'$, and $G$.
$\eta^F_\Gamma(t,\V{p})$ is a meson operator with momentum $\V{p}$ at time $t$:
\be
  \eta^F_\Gamma(t,\V{p}) = \sum_{x:x_4=t} e^{-\I \V{p}\cdot\V{x}} \,
      F_{ff'} \, \bar\psi^a_{f\alpha}(x)
      \Gamma_{\alpha\beta} \psi^a_{f'\beta}(x)
\ee
($a$ color, $f$ flavor, $\alpha$ Dirac index) with a suitably chosen Dirac
matrix $\Gamma$. In the case of the rho, $\Gamma=\gamma_j$, while for the
pion $\Gamma=\gamma_5$. A second meson operator is set at time slice $0$
with momentum $-\V{p}$, and $F'=F^\cj$. The operator ${\cal O}^G(\tau)$ has the
general form
\be
  {\cal O}^G(\tau) = \sum_{x,z,z':x_4=\tau} 
      G_{ff'} \, \bar\psi^a_{f\alpha}(z) J^{ab}_{\alpha\beta}(z,z';x)
      \psi^b_{f'\beta}(z')
\ee
where $J^{ab}_{\alpha\beta}(z,z';x)$ is a matrix that represents the flavor,
Dirac, and derivative structure of the corresponding local operator.
$x$ can be imagined as the ``center of mass'' of the operator 
while the sum over $z$ and $z'$ represents the derivative structure.

Inserting these definitions, the correlation function is
\bea &&
  \langle \eta^F_\Gamma(t,\V{p}) \, {\cal O}^G(\tau) \,
          \eta^{F'}_{\Gamma'}(0,-\V{p}) \rangle \nnm\\
  &=& V_3\, \sum_{x:x_4=\tau} \sum_{y:y_4=t} \sum_{z,z'}
      e^{-\I(\V{p}\cdot\V{y})} \,
      F_{fg} \, F'_{f'g'} \, G_{hh'} \, \Gamma_{\alpha\beta} \,
      \Gamma'_{\alpha'\beta'} \nnm\\
  &&  \times \langle
      J^{bc}_{\gamma\delta}(z,z';x) \,
      \bar\psi^a_{f\alpha}(y) \psi^a_{g\beta}(y) \,
      \bar\psi^b_{h\gamma}(z) \psi^c_{h'\delta}(z') \,
      \bar\psi^{a'}_{f'\alpha'}(0) \psi^{a'}_{g'\beta'}(0) \rangle
\eea
where $V_3$ is the volume of a time slice.

We integrate out the fermion fields in the quenched approximation
and define
\be
  \langle \psi^a_{f\alpha}(x) \bar\psi^b_{f'\beta}(y) 
  \rangle_{\mbox{\scriptsize fermions}}
  = \delta_{ff'} G^{ab}_{\alpha\beta}(U|x,y)
\ee
where $G(U|x,y)$ is the fermion propagator in the gauge field configuration
$U$ (the $U$ dependence will be indicated explicitly only when needed), and
the average is over fermion fields. There are six different contraction
terms. In four of them, two operators at the same location are contracted.
These fermion line disconnected contributions are proportional to $\tr F$,
$\tr F'$, and $\tr G$ and vanish if these matrices are chosen traceless.
However, this is in general impossible for $\tr G$, as we shall see below,
and the omission of the corresponding contraction must be regarded an
approximation, which is however consistent with quenching. We use this
approximation for the same reason we use quenching: it is very hard to go
beyond it. The remaining two terms are the fermion line connected
contributions
\bea &&
  -V_3 \sum_{x:x_4=\tau} \sum_{y:y_4=t} \sum_{z,z'}
  e^{-\I\V{p}\cdot\V{y}} \nnm\\ &&
  \Big\langle
  (\tr FGF') \, 
  \tr{}_{DC} \left(
    \Gamma G(y,z) J(z,z';x) G(z',0) \Gamma' G(0,y) \right) \nnm \\ &&
  + (\tr F'GF) \, 
  \tr{}_{DC} \left(
    \Gamma G(y,0) \Gamma' G(0,z) J(z,z';x) G(z',y) \right)
  \Big\rangle_g
\eea
where the traces are over Dirac and color indices, and the average is over 
the gauge field alone.

The two terms can be related to each other by means of the following 
relations
\bea
G(x,y)^\cj &=& \gamma_5 G(y,x) \gamma_5 \\
\gamma_5 \Gamma &=& s \Gamma^\cj \gamma_5 \\
\gamma_5 \Gamma' &=& s' {\Gamma'}^\cj \gamma_5 \\
\gamma_5 J(z,z';x)^\cj \gamma_5 &=& s_J J(z',z;x) \label{gamma5J}
\eea
where $s,s',s_J = \pm 1$, and eq.~(\ref{gamma5J}) is valid only if the
correspondig operator is suitably symmetrized in its space-time indices. For
the pion, $\Gamma=\Gamma'=\gamma_5$ and thus $s=s'=1$, while for the rho,
$\Gamma,\Gamma'\in \{\gamma_1,\gamma_2,\gamma_3\}$ and $s=s'=-1$. Then the
correlation function reduces to
\be
  -V_3 \sum_{x:x_4=\tau} \sum_{y:y_4=t}
  e^{-\I \V{p}\cdot\V{y}} \,
  \left[ (\tr FGF') M(x,y)
         + s s' s_J \, (\tr F'GF) \, M(x,y)^* \right]
\ee
with the basic single-flavor correlation function
\be \label{singflav}
  M(x,y) = \sum_{z,z'} 
      \langle \tr{}_{DC} \,\Gamma G(y,z) J(z,z';x) G(z',0) \Gamma' G(0,y)
      \rangle_g \quad.
\ee
\note{Graphic representation could go here}
Note that the calculation of this quantity on the lattice requires only two 
inversions of the fermion matrix, one at $0$ and one at $y$ or $z$.

Using the charge conjugation matrix, defined by
\be
  \gamma_\mu^T = -C^{-1} \gamma_\mu C \quad,
\ee
and the relations
\bea
  G(U|x,y) &=& C \, G(U^*|y,x)^T \, C^{-1} \\
  C \Gamma^T C^{-1} &=& \sigma \Gamma \\
  C {\Gamma'}^T C^{-1} &=& \sigma' \Gamma' \\
  C J(U^*|z',z;x)^T C^{-1} &=& 
  \sigma_J J(U|z,z';x)
\eea
(where we explicitly denoted the dependence of $J$ on the gauge field)
with $\sigma,\sigma',\sigma_J = \pm 1$ one can further show that
\be
  M(x,y)^* = \sigma \sigma' \sigma_J \, s s' s_J \, M(x,y) \quad.
\ee
We choose traceless matrices for $F$ and $F'$,
\be
  F  = \left( \begin{array}{cc} 0 & 1 \\ 0 & 0\end{array} \right)
  \quad,\qquad
  F' = \left( \begin{array}{cc} 0 & 0 \\ 1 & 0\end{array} \right)
  \quad,
\ee
and therefore
\be
  \tr F'G F = G_{11}
  \quad, \qquad
  \tr F G F' = G_{22} \quad.
\ee

We finally arrive at the following expression relating the propagators
$M(x,y)$ to the three-point correlation function:
\bea
  \noalign{\strut$\ds
  \langle \eta^F_\Gamma(t,\V{p}) \, {\cal O}^G(\tau) \,
          \eta^{F'}_{\Gamma'}(0,-\V{p}) \rangle =$}
  &=&
  -V_3 \,
  \left( G_{11} + \sigma_J \, G_{22} \, \right) \,
  \times \sum_{x:x_4=\tau} \sum_{y:y_4=t}  \,
  e^{-\I \V{p}\cdot\V{y}} \,
  M(x,y)
\eea
For an operator with $n$ derivatives, $\sigma_J$ is
$(-1)^{n+n_5+1}$, where $n_5=1$ if the operator contains a $\gamma_5$ 
matrix, $n_5=0$ otherwise. Thus, for odd $n+n_5$, $G$ must not be
traceless.

The analogous expression for the two-point correlation function reads
\bea 
  \noalign{\strut$\ds\qquad
  \langle \eta^F_\Gamma(t,\V{p}) \, \eta^{F'}_{\Gamma'}(0,-\V{p}) \rangle =$}
  &=& -V_3 \sum_{x:x_4=t} e^{-\I\V{p}\cdot\V{x}}
  \langle \tr{}_{DC} \, G(x,0) \Gamma' G(0,x) \Gamma \rangle_g \quad.
\eea


\section{Lattice and continuum operators}
\label{secLatCont}

Eqs.~(\ref{ratio}) and (\ref{ratiopion}) relate the numerically computable
ratios $R_{ij}$ and $R$ to expectation values of Euclidean lattice
operators. To connect them with the corresponding continuum Minkowski-space
operators, we first introduce the continuum matrix element of the
renormalized Euclidean operator $\hat O^{\rm cont}_{\rm r}$ by the relation
\be \label{cme}
    Z_{\hat O} \, \brakket{\V{p}\lambda}{\hat O}{\V{p}\lambda}
   = \frac{1}{2E(\V{p})} \, \frac{1}{2\kappa} \,
  {}^{\rm cont}\langle \V{p}\lambda | \hat O^{\rm cont}_{\rm r} |
        \V{p}\lambda \rangle^{\rm cont} \quad.
\ee
The factor $2E(\V{p})$ is a consequence of the different normalization on the
lattice and in the continuum:
\bea
  \braket{\V{p}\lambda}{\V{p}^{\,\prime}\lambda'} &=& 
                                              \delta_{\V{p},\V{p}^{\,\prime}} 
                                               \delta_{\lambda,\lambda'}
  \nnm\\
  {}^{\rm cont}\braket{\V{p}\lambda}{\V{p}^{\,\prime}\lambda'}^{\rm cont}
  &=& (2\pi)^3 \, 2E(\V{p}) \, \delta_{\lambda,\lambda'}  \,
      \delta(\V{p}-\V{p}^{\,\prime}) \quad,
\eea
and $2\kappa$ comes from the definition of the Wilson fermion action on the
lattice. $Z_{\hat O}$ is the renormalization constant of the operator $\hat
O$.

In the following, we shall use the renormalization constants calculated in 
one-loop lattice perturbation theory in the chiral limit \cite{Renorm}. 
They can be written in the form
\begin{equation}
  Z_{\hat O} = 1 - \frac{g^2}{16\pi^2} \, C_F \, 
   \left( \gamma_{\hat O} \ln(a\mu) + B_{\hat O} - B^c_{\hat O} \right)
    \,,
\end{equation}
where $C_F = 4/3$, $g$ denotes the bare coupling constant, and
$\mu$ is the renormalization scale. Note that here the lattice spacing $a$
has been introduced explicitly. The finite contribution $B_{\hat O}$
is fixed in the momentum subtraction renormalization scheme, whereas
$B^c_{\hat O}$ represents the contribution of the continuum operator in the
$\overline{\mbox{MS}}$ scheme with an anticommuting $\gamma_5$. 
Hence multiplication by $Z_{\hat O}$ leads from bare operators on the
lattice to the corresponding renormalized (in the 
$\overline{\mbox{MS}}$ scheme) operators in the Euclidean continuum. 
For the renormalization scale $\mu$ we choose the inverse lattice spacing 
$a^{-1}$. Taking the physical rho mass of 770~MeV as input, we obtain from 
the lattice masses extrapolated to the chiral limit the value 
$\mu=2.4\,\mbox{GeV}$.



\section{Results}
\label{secRes}

We have collected more than 500 independent configurations on a $32\times 16^3$
lattice at $\beta=6.0$ with Wilson fermions and $r=1$. Three different
hopping parameters, $\kappa = 0.1515$, $0.153$, and $0.155$ were used. 
  They correspond to quark masses of roughly 190, 130 and 70 MeV, respectively. 
As in
\cite{Nucleon}, each gauge update consisted of a single 3-hit Metropolis
sweep followed by 16 overrelation sweeps. This cycle is repeated 50 times to
generate a new configuration. The code was run on a Quadrics QH2
data-parallel computer. For completeness, the smearing technique -- Jacobi
smearing -- we use to improve the overlap of the operator with the state
is described in Appendix \ref{appSmearing}.

The calculational procedure is as follows: We calculate in each
configuration the three-point functions (\ref{singflav}) for a large set of
operators as well as the pion and rho two-point functions. 
In Appendix \ref{appOperators} we list the operators we have actually 
studied. Those without $\gamma_5$ are labeled by the pion moments $v_n$ one 
can compute from them. The expectation value of such an operator $\hat 
O_{v_n}$ in the rho is a linear combination of $a_n$ and $d_n$. The 
operators with $\gamma_5$ are labeled by the corresponding rho matrix 
elements $r_n$.

Using two values of the momentum,
namely $\V{p}=(0,0,0)$ and $\V{p}=(\frac{2\pi}{16},0,0)$, 
we can check the continuum dispersion relation of the one-particle energies
extracted from the two-point function. It is satisfied to better than 1\%,
and even for nonzero momentum we have a good projection on the ground state
pion and rho. The particle masses we have used in our subsequent analysis
are taken from Ref.\ \cite{Improved}. They are collected in Table
\ref{tabMasses}.

For the computation of the three-point functions, the locations of the
source and the sink are held fixed at $0$ and $t=T/2=16$. 
Placing the sink at $T/2$ allows us to search for a plateau equally well on
both sides of the sink. In the case of the rho, we restrict ourselves to the
3--3 and 3--2 components.

For the denominator of the ratios we employed two different procedures:
first, we took the actual value of the propagator at midpoint, and second,
we fitted the interior 24 points of the propagator to exponential
functions and used the resulting midpoint value. The second case resulted in
somewhat smaller errors at certain values of $\kappa$ and $p^1$. We quote
our results including the uncertainty from the former procedure. 
We also tried to use the conserved vector current, as
proposed by \cite{Sachrajda}, but this did not reduce our error margins.

\begin{figure}
\vspace{-3mm}
\centerline{\hfil\epsfxsize=6cm\epsfbox{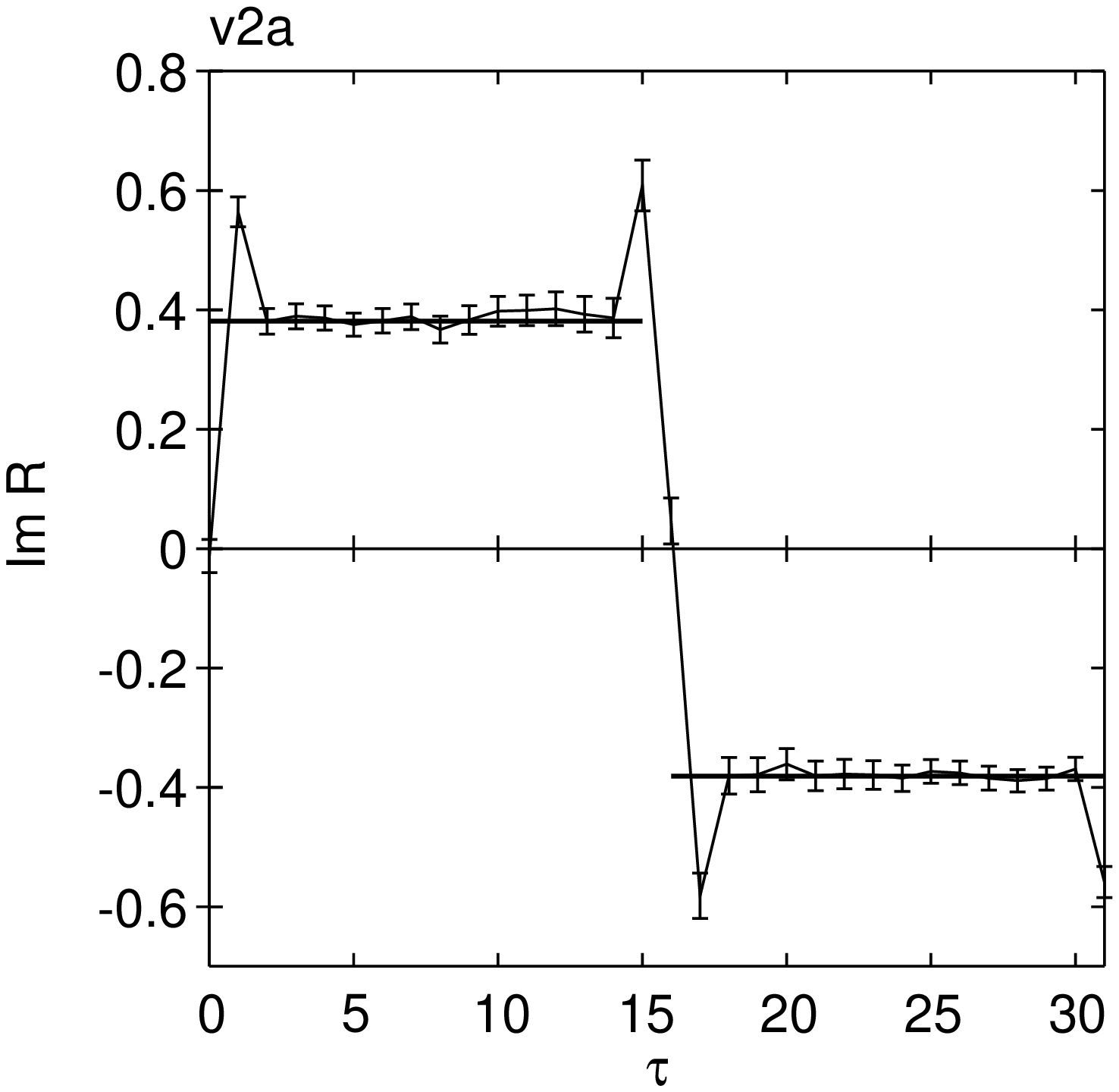}\hfil
            \epsfxsize=6cm\epsfbox{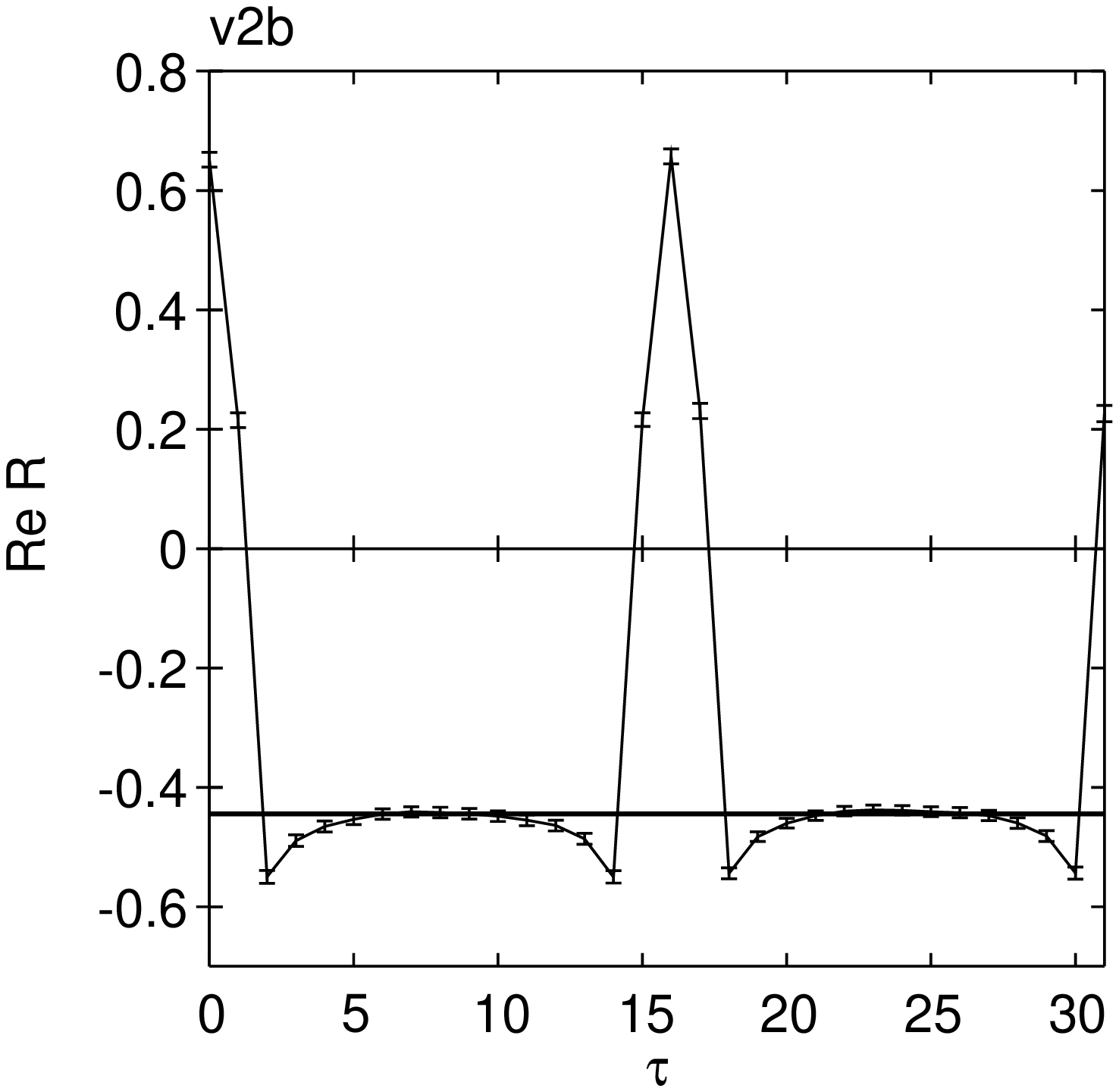}
            \hfil}
\vspace{-7mm}
\centerline{\hfil\epsfxsize=6cm\epsfbox{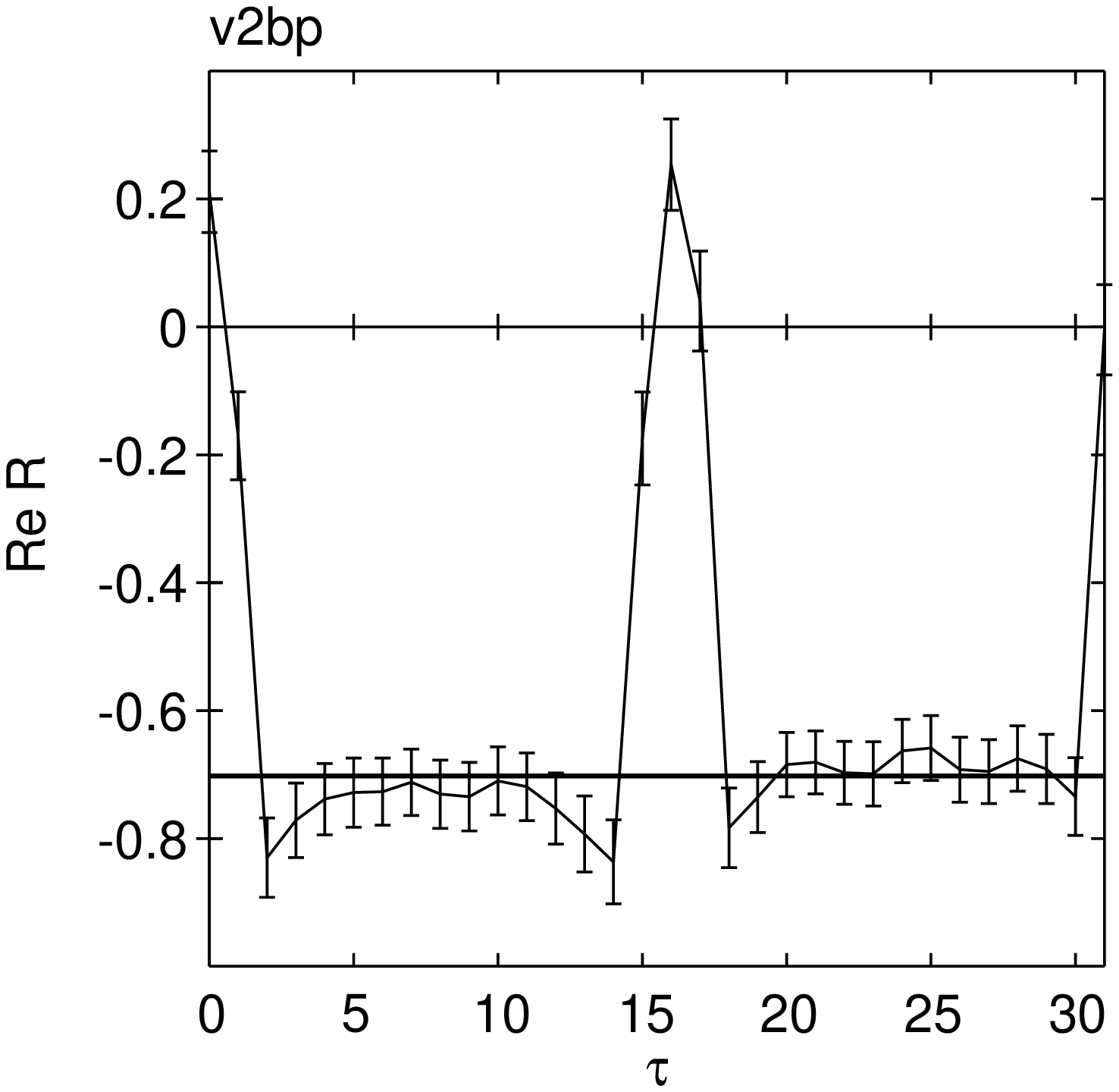}\hfil
            \epsfxsize=6cm\epsfbox{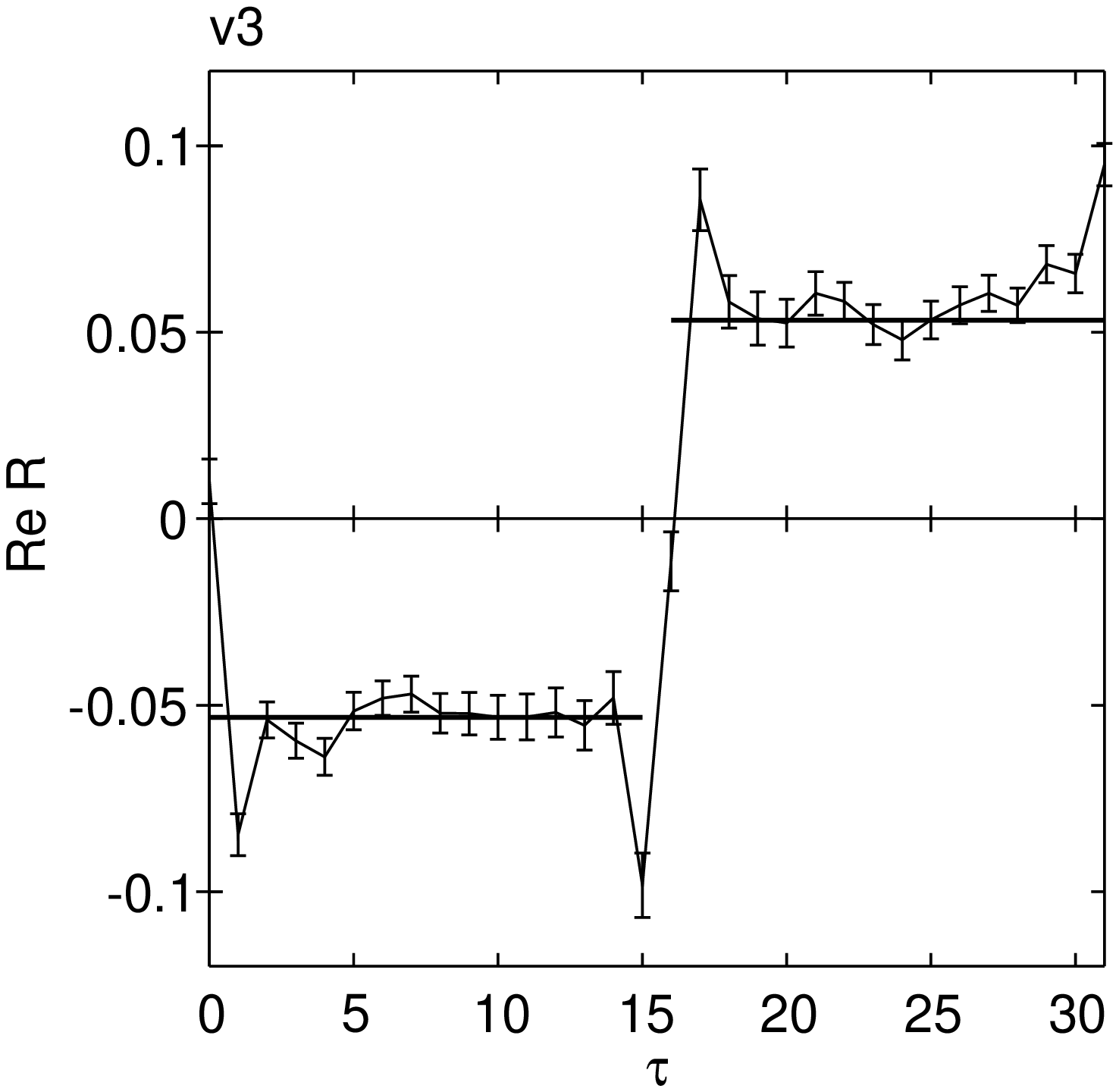}
            \hfil}
\vspace{-7mm}
\centerline{\hfil\epsfxsize=6cm\epsfbox{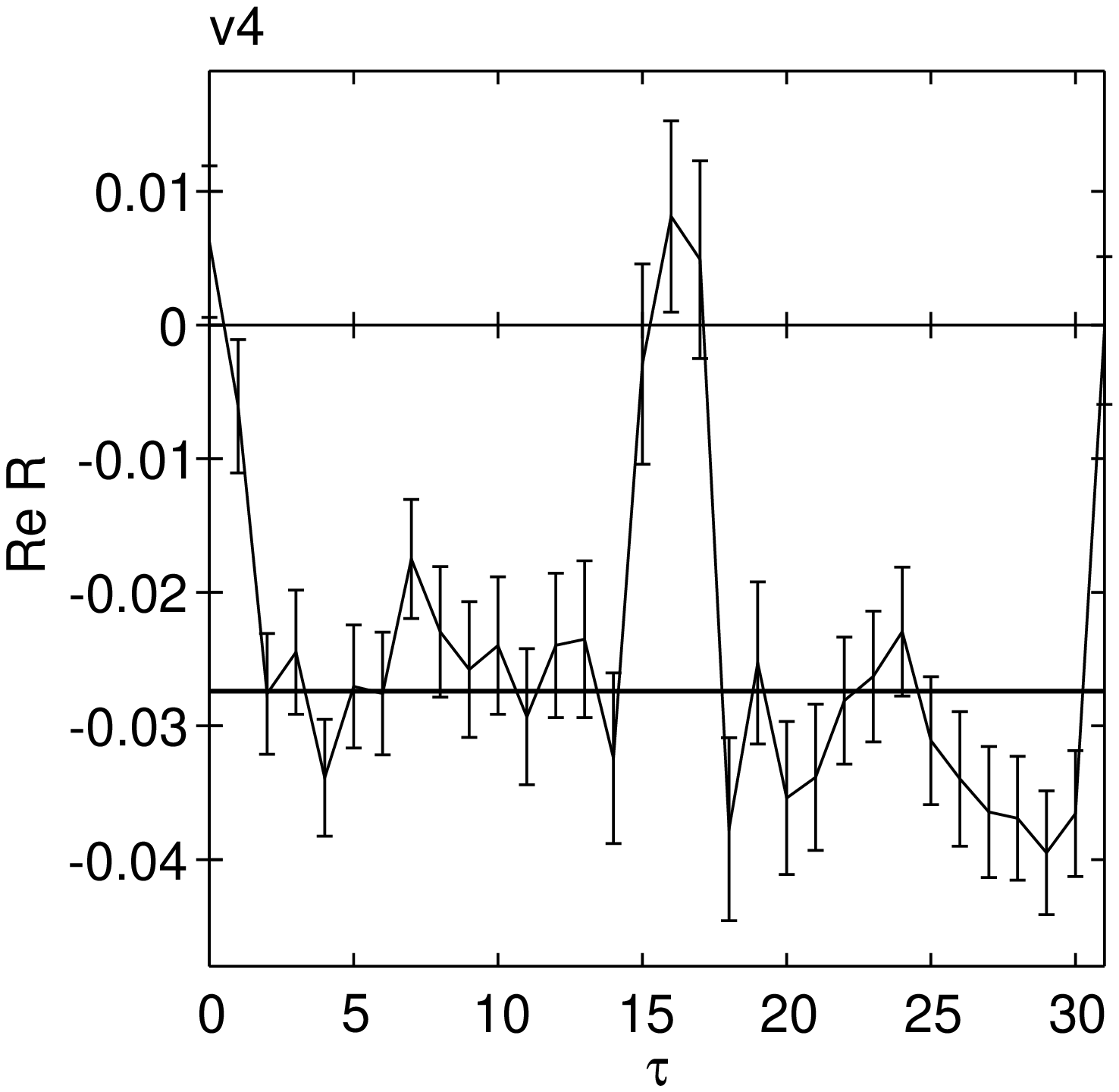}
            \hfil}
\caption{The ratio $R$ for the 
operators $\hat O_{v2a}$, $\hat O_{v2b}$, $\hat O_{v2b}$ at 
$\V{p}\ne\V{0}$,
$\hat O_{v3}$, and $\hat O_{v4}$ (left to right,
top to bottom) for the pion at $\kappa=0.153$. The horizontal line is
a fit to the central seven points on both sides.
\label{fig1}}
\end{figure}

\begin{figure}
\centerline{\hfil\epsfxsize=6cm\epsfbox{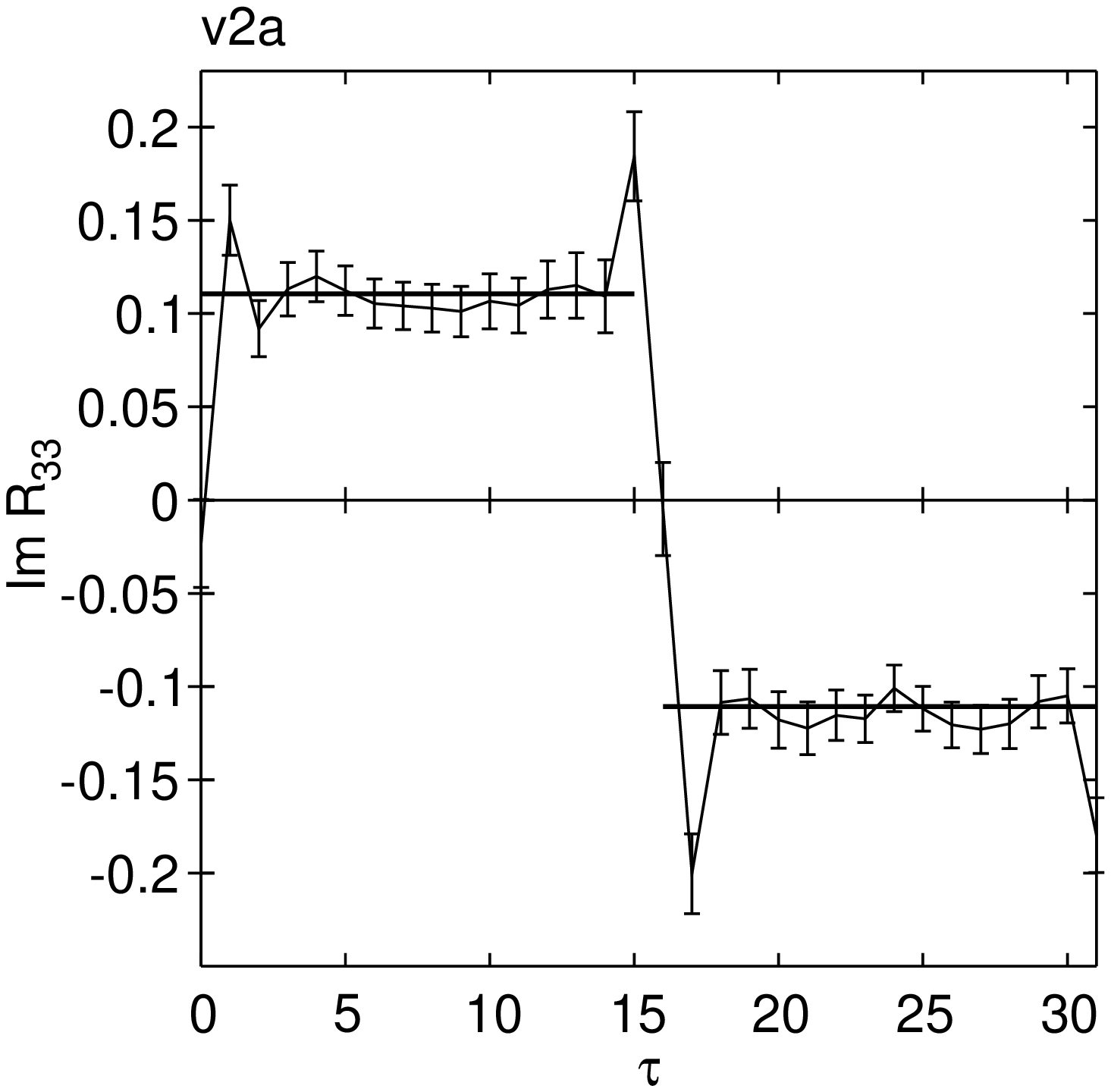}\hfil
            \epsfxsize=6cm\epsfbox{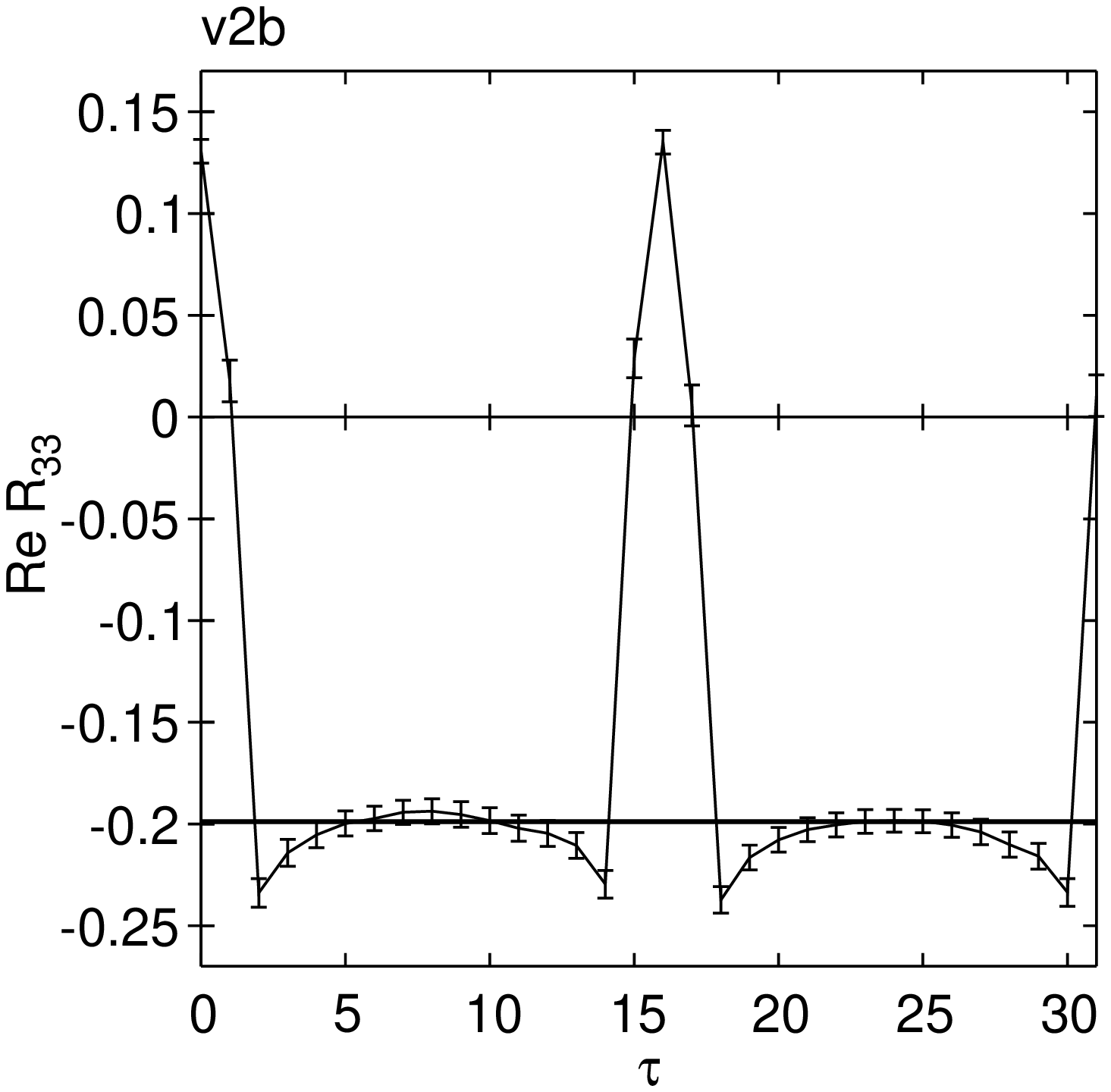}
            \hfil}
\vspace{-2mm}
\centerline{\hfil\epsfxsize=6cm\epsfbox{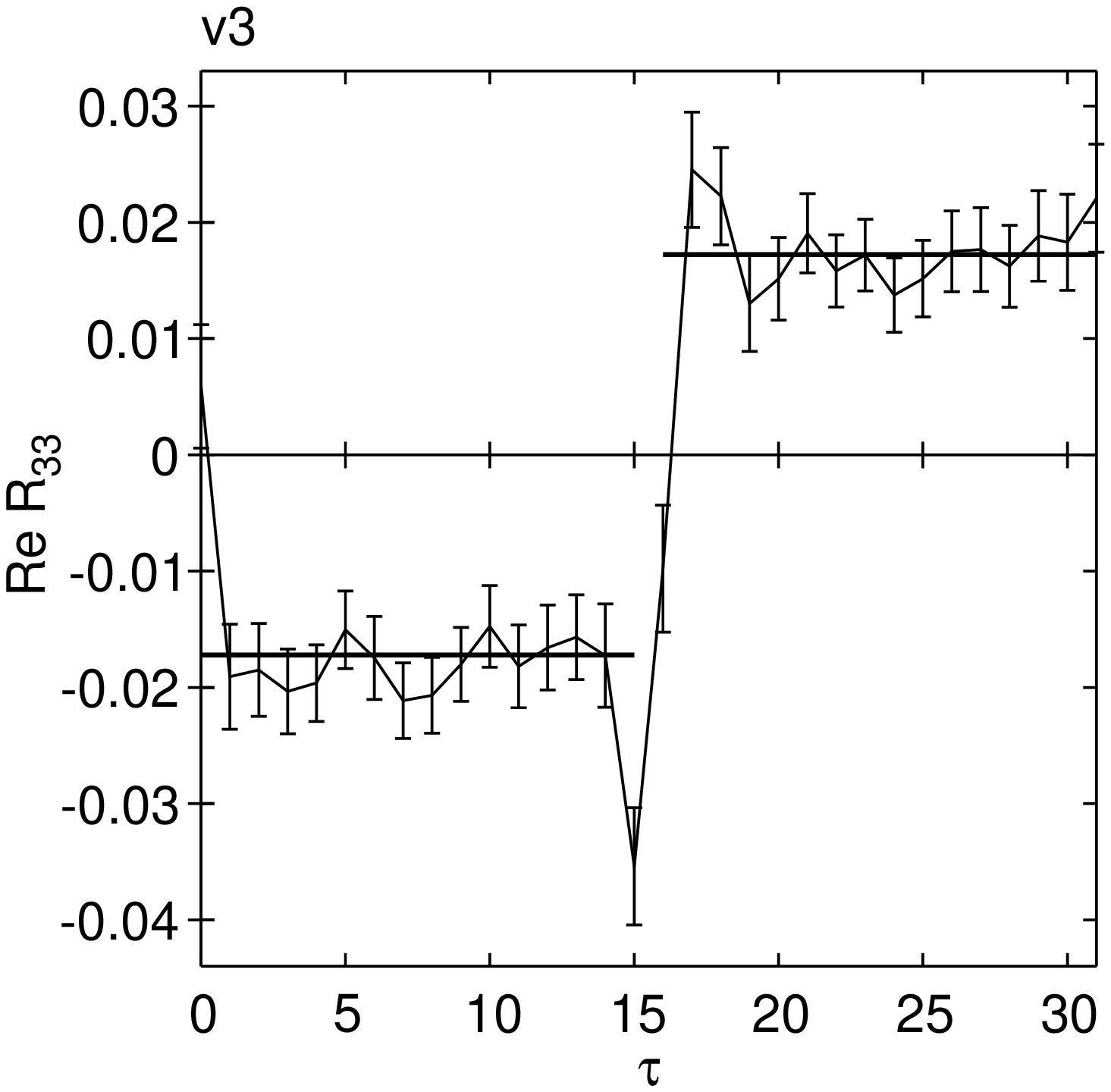}\hfil
            \epsfxsize=6cm\epsfbox{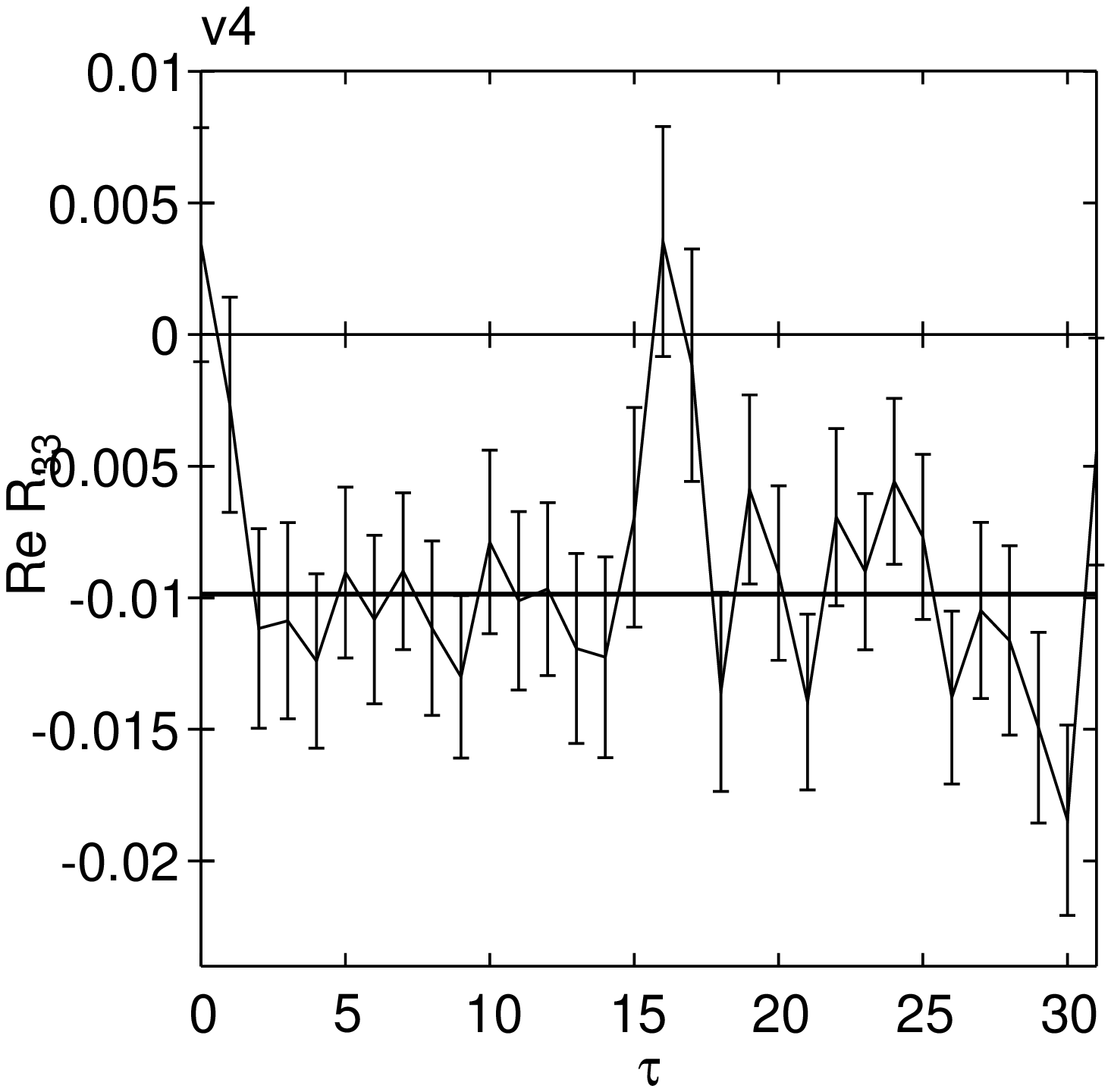}
            \hfil}
\vspace{-2mm}
\centerline{\hfil\epsfxsize=6cm\epsfbox{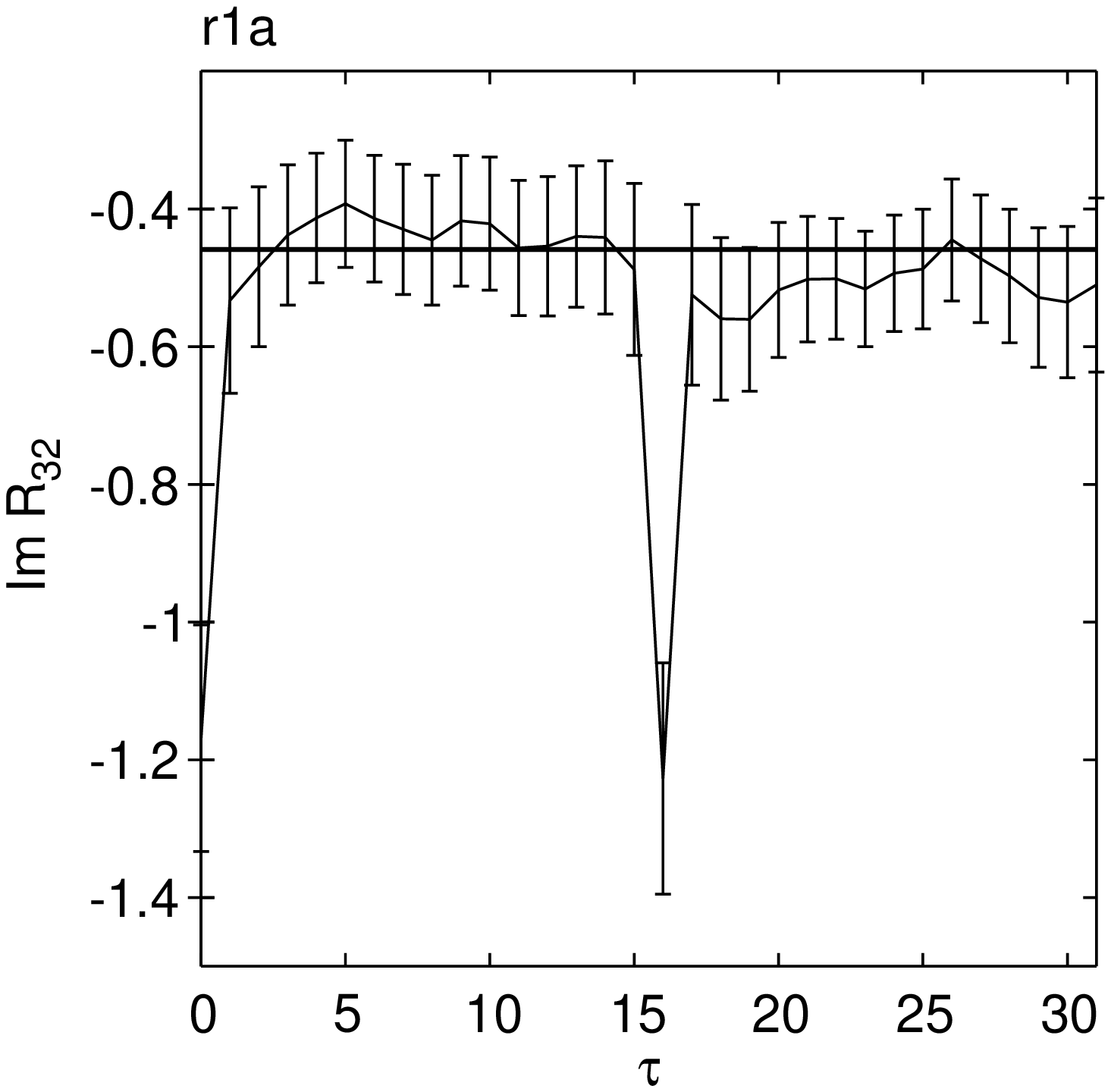}\hfil
            \epsfxsize=6cm\epsfbox{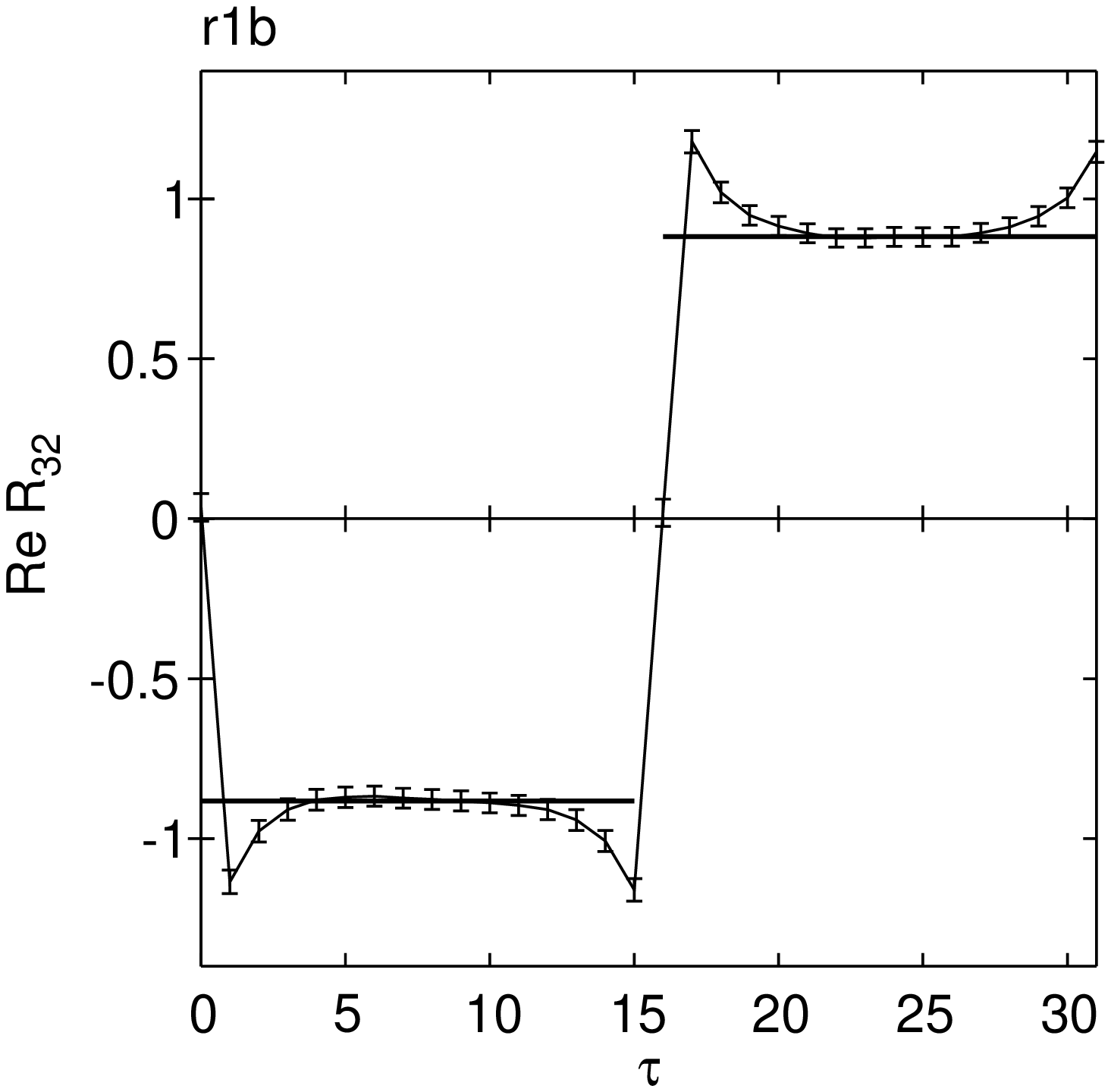}
            \hfil}
\caption{The ratios $R_{33}$ and $R_{32}$ for the 
rho operators $\hat O_{v2a}$, $\hat O_{v2b}$, $\hat O_{v3}$, $\hat O_{v4}$,
$\hat O_{r1a}$, and $\hat O_{r1b}$ (left to right, top to bottom) at
$\kappa=0.153$.
\label{fig1a}}
\end{figure}

The ratios (\ref{ratio}) and (\ref{ratiopion}) are taken as
a function of the operator insertion point
$\tau$, and a fit to the central 7 points on each side that make up the
plateau is made. The full covariance matrix
is considered in estimating the error, thus taking correlations between
neighboring points into account (in fact, only about 2 independent degrees
of freedom out of 7 survived). Some example plots are shown in
Figs.~\ref{fig1} and \ref{fig1a}.

\begin{table}
\begin{center}
\begin{tabular}{|l|c|c|c|c|}
\hline
& $\kappa=0.1515$ & $\kappa=0.153$ & $\kappa=0.155$ & 
$\kappa=\kappa_c=0.15717(3)$ \\
\hline
Pion & $0.5033(4)$ & $0.4221(4)$ & $0.2966(5)$ & 0 \\
Rho  & $0.5682(7)$ & $0.5058(8)$ & $0.4227(15)$ & 0.328(5)\\
\hline
\end{tabular}
\end{center}
\caption{Pion and rho masses in lattice units.\label{tabMasses}}
\end{table}

In a few cases we have two operators for the same reduced continuum
matrix element, distinguished by the subscripts $a$ and
$b$. $\hat{O}_{v2,a}$ and $\hat{O}_{v2,b}$ as well as 
$\hat{O}_{r2,a}$ and $\hat{O}_{r2,b}$ belong to different
representations of the hypercubic group $H(4)$. Hence the results
extracted from operator $a$ and operator $b$ have to agree only in the
continuum limit where the full $O(4)$ symmetry is hopefully restored,
and a comparison of our results obtained for finite lattice spacing
gives us some indication of the size of lattice artifacts.  
In the case of the operators associated to $v_{2,b}$, $r_{1,b}$,
and $r_{2,a}$ we denote the results obtained with nonzero momentum by
an additional subscript $p$.

\begin{figure}
\centerline{\epsfbox{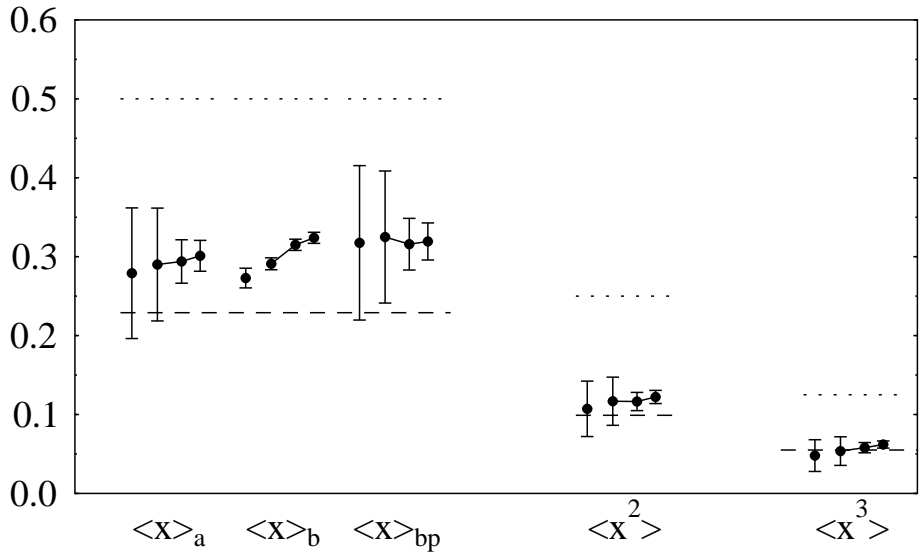}}
\caption{Estimates of the pion moments $v_n=\langle x^{n-1}\rangle$
for a single flavor. 
For each matrix element, the results from the three different $\kappa$
values are shown versus $1/\kappa$ with $\kappa$ decreasing (i.e.~with the
quark mass increasing) from left to right. The leftmost value is 
the chiral extrapolation obtained
from a linear fit. The dotted lines give the
free-field (heavy quark) limits. The dashed lines are phenomenological
valence quark values from \protect\cite{Sutton}, evaluated at 
$\mu=2.4\,\mbox{GeV}$. \label{fig2}}
\end{figure}

Applying (\ref{ratiopion}), (\ref{cme}), (\ref{Mops}), and (\ref{operespion})
we have calculated estimates for the pion moments $v_n$ from the
measured ratios $R$. The results are summarized in
Fig.~\ref{fig2} and in Table \ref{tab1}. The agreement (within errors)
of $v_{2,a}$ and $v_{2,b}$ indicates that -- at least in this case --
lattice artifacts are not too large. Assuming a linear dependence on
$1/\kappa$, i.e.\ on the bare quark mass, the values have been
extrapolated to the chiral limit $\kappa = \kappa_c = 0.15717(3)$. Since the
quark masses in our simulation are rather large 
($>$ 70MeV) we need this
extrapolation in order to obtain numbers that can sensibly be compared
with phenomenological valence quark distributions. Note, however, that the
quark mass dependence of the results is not very pronounced. Only $v_{2,b}$
shows a significant trend towards smaller values as the chiral limit is
approached which is the expected behavior. 

\begin{figure}
\vskip-4mm
\centerline{\epsfbox{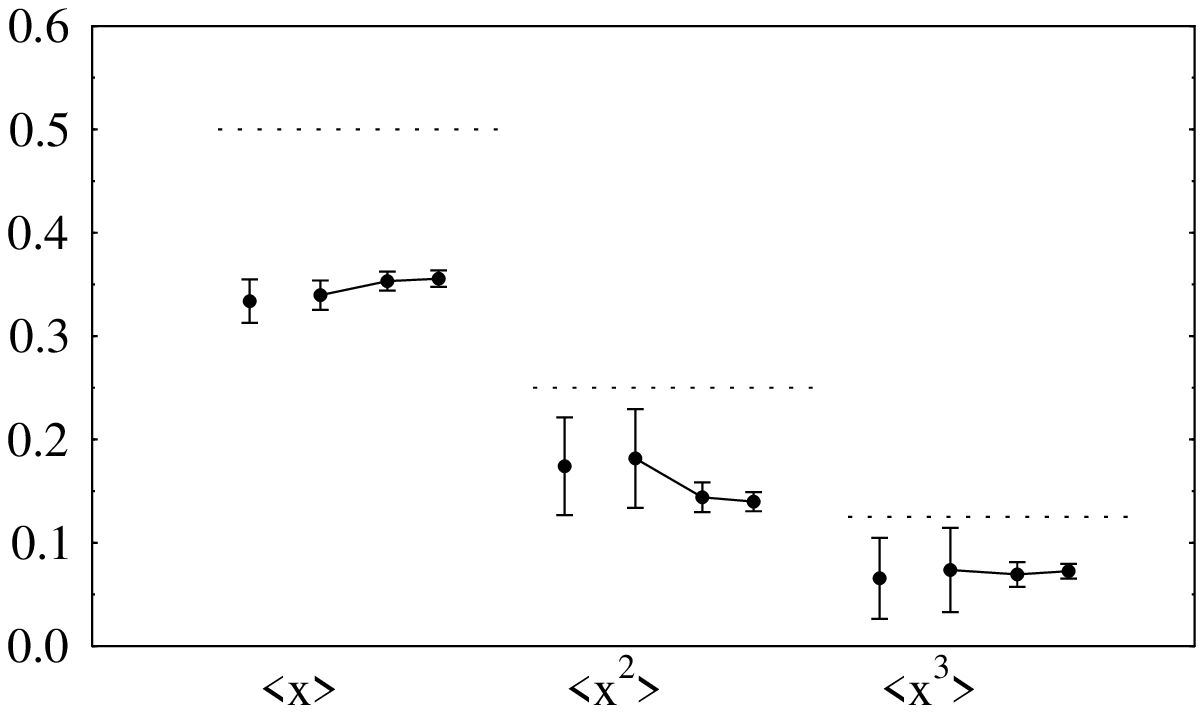}}
\vskip-8mm
\centerline{\epsfbox{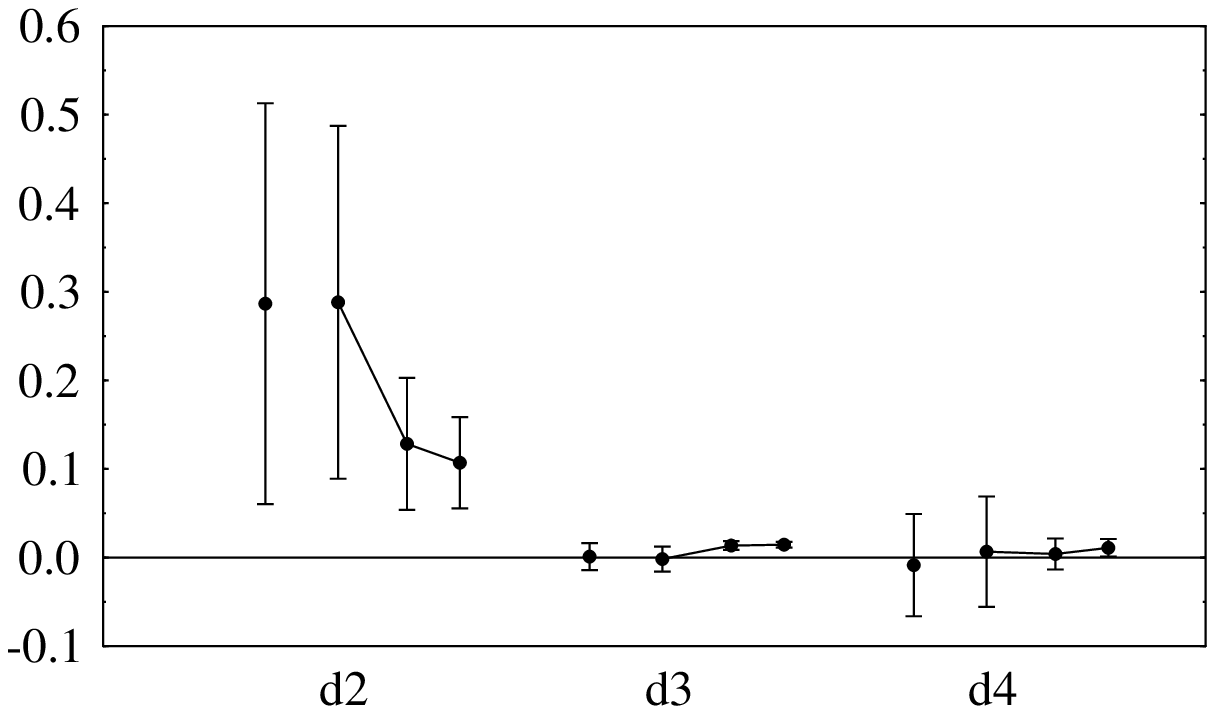}}
\vskip-8mm
\caption{Estimates for the rho moments $a_n=\langle x^{n-1}\rangle$
and $d_n$. The presentation of the data is
the same as in Fig.~\ref{fig2}.
\label{fig3}}
\end{figure}

\begin{figure}
\vskip-4mm
\centerline{\epsfbox{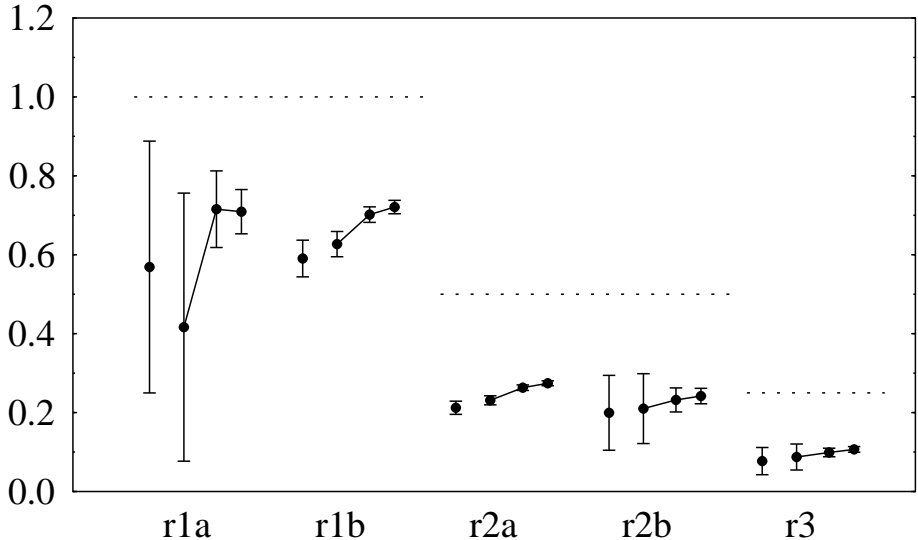}}
\vskip-8mm
\caption{Estimates for the rho moments $r_n$. The presentation of the data
is the same as in Fig.~\ref{fig2}. 
\label{fig4}}
\end{figure}

We now come to the rho results (see Figs.\ \ref{fig3}, \ref{fig4} and Table
\ref{tab1}). By means of (\ref{sign2}), (\ref{sign3}), (\ref{cme}), and
(\ref{Mops}) we pass from the ratios (\ref{ratio2}) to matrix elements whose
relation to $a_n$, $d_n$, and $r_n$ is listed in Eq.\ (\ref{operes}). The
extrapolation to the chiral limit is performed as for the pion. In the case
of the operators without $\gamma_5$, i.e.~those labeled by $v_n$,
we encounter the problem that instead of
one number ($v_n$) we have to extract the two quantities $a_n$ and $d_n$
from the matrix elements. 

Therefore we proceed as follows: The expectation value of $\hat O_{v2,b}$ 
at $\V{p} = \V{0}$ gives us directly $a_2$, and $d_2$ can then be 
calculated from $\hat O_{v2,a}$   and $\hat O_{v2,b}$ 
at nonvanishing momentum. The expectation 
values of $\hat O_{v3}$ and $\hat O_{v4}$, on the other hand, are 
proportional to $d_3$ and $d_4$, respectively, if $\V{p} = \V{0}$. With 
$d_3$ and $d_4$ computed from these matrix elements we use the 
corresponding results for 
$\V{p} \ne \V{0}$ to calculate $a_3$ and $a_4$.

{}From the matrix elements of the operators with $\gamma_5$ we can easily
extract $r_1$, $r_2$, and $r_3$. The estimates $r_{2,a}$, $r_{2,ap}$,
and $r_{2,b}$ for $r_2$ agree within the errors. Thus also in this
case we do not observe significant discretization effects. 

\begin{table}
\begin{tabular}{|l|c|c|c|c|}
\hline
&$\kappa=0.1515$ & $\kappa=0.153$ & $\kappa=0.155$ & $\kappa=\kappa_c=0.15717$\\ \hline
\multicolumn{4}{l}{PION}\\ \hline
$v_{2,a}=\langle x\rangle_a$ & $0.301(20)       $ & $0.294(28)       $ & $0.290(71)       $ & $0.279(83)       $\\
$v_{2,b}=\langle x\rangle_b$ & $0.3239(70)      $ & $0.3150(71)      $ & $0.2910(75)      $ & $0.273(12)       $\\
$v_{2,bp}=\langle x\rangle_{bp}$ & $0.319(23)       $ & $0.316(33)       $ & $0.325(84)       $ & $0.318(98)       $\\
$v_3=\langle x^2\rangle$ & $0.1222(83)      $ & $0.116(12)       $ & $0.117(31)       $ & $0.107(35)       $\\
$v_4=\langle x^3\rangle$ & $0.0619(45)      $ & $0.0580(65)      $ & $0.054(18)       $ & $0.048(20)       $\\
\hline                                                                    
\multicolumn{4}{l}{RHO}\\ \hline                                          
$a_2=\langle x\rangle$ & $0.3555(80)      $ & $0.3531(93)      $ & $0.340(14)       $ & $0.334(21)       $\\
$a_3=\langle x^2\rangle$ & $0.1398(93)      $ & $0.144(14)       $ & $0.182(48)       $ & $0.174(47)       $\\
$a_4=\langle x^3\rangle$ & $0.0725(72)      $ & $0.069(12)       $ & $0.074(41)       $ & $0.066(39)       $\\
\hline
$d_2$           & $0.107(52)       $ & $0.128(75)       $ & $0.29(20)        $ & $0.29(23)        $\\
$d_3$           & $0.0145(32)      $ & $0.0135(49)      $ & $-0.002(14)      $ & $0.001(15)       $\\
$d_4$           & $0.0109(100)     $ & $0.004(17)       $ & $0.007(62)       $ & $-0.009(58)      $\\
\hline
$r_{1,a}$       & $0.709(56)       $ & $0.715(97)       $ & $0.42(34)        $ & $0.57(32)        $\\
$r_{1,b}$       & $0.721(17)       $ & $0.702(20)       $ & $0.627(32)       $ & $0.590(46)       $\\
$r_{1,bp}$      & $0.680(56)       $ & $0.62(13)        $ & $0.32(44)        $ & $0.33(42)        $\\
$r_{2,a}$       & $0.2743(62)      $ & $0.2631(70)      $ & $0.231(12)       $ & $0.212(17)       $\\
$r_{2,ap}$      & $0.257(17)       $ & $0.243(25)       $ & $0.216(69)       $ & $0.198(76)       $\\
$r_{2,b}$       & $0.242(20)       $ & $0.232(30)       $ & $0.210(89)       $ & $0.199(95)       $\\
$r_3$           & $0.1067(71)      $ & $0.099(11)       $ & $0.087(33)       $ & $0.077(34)       $\\
\hline
\end{tabular}
\caption{Result overview for a single flavor.
The numbers refer to the $\overline{\mbox{MS}}$ scheme with a 
renormalization scale $\mu\approx 2.4\,\mbox{GeV}$. The last column gives 
the result of the extrapolation to the chiral limit.
\label{tab1}}
\end{table}

 \section{Discussion}
\label{secConc}

We have calculated the lowest three moments of the structure functions of the 
pion and the rho meson, restricting 
ourselves to the leading twist-2 
operators in the operator product expansion.

For the pion, we can compare our numbers in the chiral limit with 
the experimental data~\cite{Sutton}. Our result for 
$\langle x \rangle$ is larger than phenomenology suggests. This is to
be expected as our quenched lattice calculation does not contain any
sea quarks and the valence quarks will therefore carry more of the
momentum. The results for $\langle x^2 \rangle$ and 
$\langle x^3 \rangle$, on the other hand, are consistent with the
phenomenological numbers.
Our results also agree with the early lattice 
calculations of Martinelli and Sachrajda~\cite{Sachrajda} as well as with
various model calculations~\cite{model}.

The unpolarized rho structure function looks very similar to the pion 
structure function, at least for the quark masses that we have considered. 
In the pion the quarks carry about 60\% of the total momentum, while 
in the rho they carry about 70\% at the smallest quark mass. The higher 
moments are in agreement with each other within the error bars. Thus the 
assumption $F_1^\rho(x) \sim F_1^\pi(x)$ often used in 
phenomenological estimates may well be justified. 

The lowest moment $r_1$ of the polarized structure function $g_1$ indicates 
that the valence quarks carry about 60\% of the total spin of the rho.
For comparison, a similar quenched calculation for the nucleon gave a
quark spin fraction of about the same value~\cite{Nucleon}, which is 
reduced to 18\% by sea quark contributions~\cite{seaquark}. It is very
likely that the same will also happen here.

The structure functions $b_1, b_2$ measure the difference in quark 
distributions of a (spin projected) $m = 1$ and $m = 0$ rho meson. If the 
quarks were in a relative $s$-wave state in the infinite momentum frame, we
would expect $b_1, b_2$ to be zero. The lowest moment $d_2$ turns out to
be positive and surprisingly large on the scale of $a_2$, albeit with large
statistical errors. Perhaps this indicates that the valence quarks have a 
substantial orbital angular momentum. This could also explain a relatively
small quark spin fraction.

\section*{Acknowledgements}
This work was supported in part by the Deutsche Forschungsgemeinschaft. The
numerical calculations were performed on the Quadrics parallel computer at
Bielefeld University. We wish to thank the computer center for its help.
C.B.~thanks the   Studienstiftung des Deutschen Volkes  for its support.


\appendix

\section{Conventions}
\label{appConv}

The Minkowski space metric has the signature $(1,-1,-1,-1)$. Minkowski 
and Euclidean components are related by
\be
  x_4 = \I x^{(M)0} = \I x^{(M)}_0 \quad, \qquad
  x_j = x^{(M)j} = - x^{(M)}_j \quad,
\ee
where $j$ refers to spatial indices. Unless explicitly mentioned, we label
Minkowski-space variables by an $(M)$. 

Covariant derivatives are defined in Minkowski space as
\be
  D^{(M)\mu} = \partial^{(M)\mu} -\I g A^{(M)\mu} \\
\ee
and are related to their Euclidean counterparts as follows:
\be
  \I D^{(M)0} = -D_4 \quad,\qquad
  \I D^{(M)j} = -\I D_j \quad,
\ee
similarly for the gamma matrices:
\be
  \gamma^{(M)0} = \gamma_4 \quad,\qquad
  \gamma^{(M)j} = \I \gamma_j \quad.
\ee
The $\gamma_5$ matrix is defined as
\bea
  \gamma_5^{(M)} &=& \I \gamma^{(M)0} \gamma^{(M)1} \gamma^{(M)2} \gamma^{(M)3}
  \nnm\\
  \gamma_5 &=& \gamma_1 \gamma_2 \gamma_3 \gamma_4
           = -\gamma_5^{(M)} \quad.  
\eea

The momentum of the particles is chosen in the $1$-direction, 
$\V{p} = (p,0,0)$. Polarization vectors for vector particles satisfy
\be
  p^{(M)\mu} \epsilon^{(M)}_\mu(\V{p},\lambda) = 0 \quad,\qquad
  \epsilon^{(M)\mu}(\V{p},\lambda) \epsilon^{(M)}_\mu(\V{p},\lambda') = -m^2 
  \, \delta_{\lambda\lambda'}
\ee
($\lambda=\pm,0$)
and have the explicit form
\be
  \epsilon^{(M)\mu}(\V{p},\lambda) =
  \left( \frac{\V{p}\cdot\V{e}_\lambda}{m},
         \V{e}_\lambda + \frac{\V{p}\cdot\V{e}_\lambda}{m(m+E)} \V{p}
  \right)
\ee
with the basis vectors
\begin{eqnarray}
 \V{e}_\pm &=& \mp\frac{m}{\sqrt{2}} \, (0,1,\pm \I) \\
 \V{e}_0 &=& m \, (1,0,0)  \quad.
\end{eqnarray}
They satisfy the completeness relation
\be \label{epsilonsum}
  \sum_\lambda \epsilon^{(M)*}_i(\V{p},\lambda) \epsilon^{(M)}_j(\V{p},\lambda)
  = m^2 \left( \delta_{ij} + \frac{1}{m^2} p_i p_j \right) \quad.
\ee
Note that in Euclidean space $\epsilon_j(\V{p},\lambda) = 
\epsilon^{(M)j}(\V{p},\lambda)$.

\section{Operators}
\label{appOperators}

On the lattice, the choice of the operators to look at is a nontrivial
matter, because the discretization reduces the symmetry group of
(Euclidean) space-time from $O(4)$ to the hypercubic group 
$H(4) \subset O(4)$. 
Hence the lattice operators have to be classified according to
$H(4)$ and one should choose operators belonging to a definite
irreducible representation of $H(4)$.
Since $H(4)$ is a finite group, the restrictions imposed
by symmetry are less stringent than in the continuum and the
possibilities for mixing increase. 
Whereas mixing with operators of the same dimension is supposed to
be treatable by perturbation theory, the mixing coefficients for
lower-dimensional operators have to be calculated nonperturbatively.
Hence one would like to avoid mixing with lower-dimensional 
operators whenever possible.
On the other hand, as the spin grows, operators with no
mixing at all require more and more nonvanishing momentum 
components in the calculation of their forward hadronic matrix 
elements, which makes their Monte Carlo evaluation increasingly 
difficult. So some kind of compromise is needed. 

Due to our use of the quenched approximation 
purely gluonic operators cannot mix with two-quark
operators and we may restrict ourselves to the latter.
Guided by their $H(4)$ classification given in
Ref.~\cite{Groupth} we have chosen the following operators in Euclidean
space: 
\begin{eqnarray}
  \hat{O}_{v2,a} & = & \hat{O}_{\{41\}} \quad,\nonumber \\
  \hat{O}_{v2,b} & = & \hat{O}_{44} - \frac{1}{3} \left(\hat{O}_{11} +
                       \hat{O}_{22} + \hat{O}_{33} \right)
                                                        \quad,\nonumber \\
  \hat{O}_{v3} & = & \hat{O}_{\{114\}} - \frac{1}{2} 
                    \left(\hat{O}_{\{224\}} + \hat{O}_{\{334\}} \right)
                                                        \quad,\nonumber \\
  \hat{O}_{v4}   & = & \hat{O}_{\{1122\}} + \hat{O}_{\{3344\}} +
                       \hat{O}_{\{1133\}} + \hat{O}_{\{2244\}} -
                      2 \hat{O}_{\{1144\}} -2 \hat{O}_{\{2233\}} 
                                                        \quad,\nonumber \\
  \hat{O}_{r1,a} & = & \hat{O}^5_4                      \quad,\nonumber \\
  \hat{O}_{r1,b} & = & \hat{O}^5_1                      \quad,\nonumber \\
  \hat{O}_{r2,a} & = & \hat{O}^5_{\{41\}} \quad,\nonumber \\
  \hat{O}_{r2,b} & = & \hat{O}^5_{44} - \frac{1}{3} \left(\hat{O}^5_{11} +
                       \hat{O}^5_{22} + \hat{O}^5_{33} \right)
                                                        \quad,\nonumber \\
  \hat{O}_{r3} & = & \hat{O}^5_{\{114\}} - \frac{1}{2} 
                    \left(\hat{O}^5_{\{224\}} + \hat{O}^5_{\{334\}} \right)
                    \quad.
\end{eqnarray}
For $v_2$ and $r_2$ we have two operators, which belong to the same
$O(4)$ multiplet in the continuum limit but transform according to
inequivalent representations of $H(4)$. Hence their matrix elements
provide a test for the restoration of $O(4)$ symmetry. 
The renormalization constants
for these operators in the $\overline{\mbox{MS}}$ scheme
are listed in Table \ref{tabZ}.

\begin{table}
\hbox to\hsize{\hfill
\begin{tabular}{|c|c|}
\hline
Operator $\hat O$ & $Z_{\hat O}$ \\
\hline
$\hat{O}_{v2,a}$ & 0.989196 \\
$\hat{O}_{v2,b}$ & 0.978369 \\
$\hat{O}_{v3}$   & 1.102397 \\
$\hat{O}_{v4}$ & 1.229911 \\
\hline
\end{tabular}
\hfill
\begin{tabular}{|c|c|}
\hline
Operator $\hat O$ & $Z_{\hat O}$ \\
\hline
$\hat{O}_{r1,a}$, $\hat{O}_{r1,b}$ & 0.866625 \\
$\hat{O}_{r2,a}$ & 0.997086 \\
$\hat{O}_{r2,b}$ & 0.998587 \\
$\hat{O}_{r3}$   & 1.108573 \\
\hline
\end{tabular}
\hfill}
\caption{Renormalization constants $Z$. \label{tabZ}}
\end{table}

Concerning the mixing properties a few remarks are in order. 
Mixing with operators of equal or lower dimension is excluded for the
operators $\hat{O}_{v2,a}$, $\hat{O}_{v2,b}$, $\hat{O}_{r1,a}$, 
$\hat{O}_{r1,b}$, $\hat{O}_{r2,a}$, $\hat{O}_{r2,b}$. 
The case of the operator
$\hat{O}_{v3}$, for which there are two further operators with the same
dimension and the same transformation behavior, is discussed in 
ref.~\cite{Groupth}. The operators
$\hat{O}_{v4}$ and $\hat{O}_{r3}$, on the other hand, could in
principle mix not only with operators of the same dimension but also with an
operator of one dimension less and different chiral properties. It is of the
type
\begin{equation}
 \bar{\psi} \sigma_{\mu \nu} \gamma_5
 \Dd{\mu_1} \Dd{\mu_2} \cdots \Dd{\mu_n} \psi \,,
\end{equation}
where $n=2$ in the case of $\hat{O}_{v4}$,
and $n=1$ for $\hat{O}_{r3}$.

Our analysis ignores mixing completely. This seems to be well
justified for $\hat{O}_{v3}$. Here a perturbative calculation gives a rather
small mixing coefficient for one of the mixing operators, whereas the
other candidate for mixing does not appear at all in a 1-loop
calculation, because its Born term vanishes in forward matrix elements. 
The same is true for all operators of dimension less than or equal to six 
which transform identically to $\hat O_{v4}$: Their Born term vanishes in 
forward matrix elements, hence they do not show up in a $1$--loop 
calculation. In the case of $\hat O_{r3}$, however, the mixing is already 
visible at the $1$--loop level. The results for $v_4$ and $r_3$ have 
therefore to be considered with some caution.
 
The corresponding Minkowski operators are found by applying
eq.~(\ref{euclo}). Defining the Minkowski analogs of our Euclidean operators
by
\begin{equation} \label{Mops}
  \begin{array}{rclcrcl}
   \hat{O}_{v2,a} & = & {\rm i}\hat{O}^{(M)}_{v2,a}\quad, & \hspace{1cm} &
          \hat{O}_{v2,b} & = & - \hat{O}^{(M)}_{v2,b}  \quad,\\ 
   \hat{O}_{v3} & = & - \hat{O}^{(M)}_{v3}\quad, & {} &
          \hat{O}_{v4} & = & \hat{O}^{(M)}_{v4}\quad, \\
   \hat{O}_{r1,a} & = & - \hat{O}^{(M)}_{r1,a}\quad,  & {} &
          \hat{O}_{r1,b} & = & {\rm i}\hat{O}^{(M)}_{r1,b}\quad, \\
   \hat{O}_{r2,a} & = & -{\rm i} \hat{O}^{(M)}_{r2,a}\quad, & {} &
          \hat{O}_{r2,b} & = & \hat{O}^{(M)}_{r2,b}\quad, \\
   \hat{O}_{r3} & = & \hat{O}^{(M)}_{r3}\quad,
  \end{array}
\end{equation}
we have
\begin{eqnarray}
  \hat{O}^{(M)}_{v2,a} & = & \hat{O}^{(M)\{01\}} \quad,\nonumber \\
  \hat{O}^{(M)}_{v2,b} & = & \hat{O}^{(M)00} + 
                       \frac{1}{3} \left(\hat{O}^{(M)11} +
                       \hat{O}^{(M)22} + \hat{O}^{(M)33} \right)
                                                        \quad,\nonumber \\
  \hat{O}^{(M)}_{v3} & = & \hat{O}^{(M)\{110\}} - \frac{1}{2} 
                    \left(\hat{O}^{(M)\{220\}} + \hat{O}^{(M)\{330\}} \right)
                                                        \quad,\nonumber \\
  \hat{O}^{(M)}_{v4} & = & 
                {} - \hat{O}^{(M)\{1122\}} + \hat{O}^{(M)\{3300\}} -
        \hat{O}^{(M)\{1133\}} + \hat{O}^{(M)\{2200\}}   \nonumber \\
  {} &{}&       {} -  2 \hat{O}^{(M)\{1100\}} +2 \hat{O}^{(M)\{2233\}} 
                                                        \quad,\nonumber \\
  \hat{O}^{(M)}_{r1,a} & = & \hat{O}_5^{(M)0}                \quad,\nonumber \\
  \hat{O}^{(M)}_{r1,b} & = & \hat{O}_5^{(M)1}            \quad,\nonumber \\
  \hat{O}^{(M)}_{r2,a} & = & \hat{O}_5^{(M)\{01\}} \quad,\nonumber \\
  \hat{O}^{(M)}_{r2,b} & = & \hat{O}_5^{(M)00} + 
                       \frac{1}{3} \left(\hat{O}_5^{(M)11} +
                       \hat{O}_5^{(M)22} + \hat{O}_5^{(M)33} \right)
                                                        \quad,\nonumber \\
  \hat{O}^{(M)}_{r3} & = & \hat{O}_5^{(M)\{110\}} - \frac{1}{2} 
               \left(\hat{O}_5^{(M)\{220\}} + \hat{O}_5^{(M)\{330\}} \right)
  \quad.
\end{eqnarray}

We can now use (\ref{opepion}), (\ref{hoodbhoy}), and (\ref{hoodbhoy2}) to
calculate the expectation values of the Minkowski space operators. For the
pion, one obtains with 
$p^2 = p^3 = 0$
\begin{eqnarray} \label{operespion}
 \langle \hat{O}^{(M)}_{v2,a} \rangle & = & 2 v_2 p^0 p^1      \quad,\nonumber \\
 \langle \hat{O}^{(M)}_{v2,b} \rangle & = & 2 v_2 \left( (p^0)^2
                                  + \frac{1}{3} (p^1)^2 \right)  \quad,\nonumber \\
 \langle \hat{O}^{(M)}_{v3}   \rangle & = & 2 v_3 (p^1)^2 p^0
                                                        \quad,\nonumber \\
 \langle \hat{O}^{(M)}_{v4}   \rangle & = & -4 v_4 (p^1)^2 (p^0)^2 \quad.
\end{eqnarray}
For the rho with polarization $\lambda = \pm$, one finds
\begin{eqnarray}
 \langle \hat{O}^{(M)}_{v2,a} \rangle & = & 
                 2 \left( a_2 - \frac{1}{3} d_2 \right) p^0 p^1
                                                       \quad,\nonumber \\ 
 \langle \hat{O}^{(M)}_{v2,b} \rangle & = & 
                   2 \left( a_2 - \frac{1}{3} d_2 \right)
                 \left( (p^0)^2 + \frac{1}{3} (p^1)^2 \right)  
                  + \frac{2}{3} m^2 d_2   \quad,\nonumber \\
 \langle \hat{O}^{(M)}_{v2,b} \rangle_{p^1=0} & = & 2 a_2 m^2
                                                       \quad,\nonumber \\ 
 \langle \hat{O}^{(M)}_{v3}   \rangle & = & 
                  2 \left( a_3 - \frac{1}{3} d_3 \right)
                   (p^1)^2 p^0 - \frac{1}{3} m^2 d_3 p^0
                                                        \quad,\nonumber \\
 \langle \hat{O}^{(M)}_{v4} \rangle & = & 
           -4 \left( a_4 - \frac{1}{3} d_4 \right) (p^1)^2 (p^0)^2
            + \frac{1}{3} m^4 d_4                        \quad,\nonumber \\
 \langle \hat{O}^{(M)}_{r1,a} \rangle & = & \pm 2 r_1 p^1 \quad,\nonumber \\
 \langle \hat{O}^{(M)}_{r1,b} \rangle & = & \pm 2 r_1 p^0 \quad,\nonumber \\
 \langle \hat{O}^{(M)}_{r2,a} \rangle & = & 
                \pm r_2 \left( (p^0)^2 + (p^1)^2 \right)  \quad,\nonumber \\
 \langle \hat{O}^{(M)}_{r2,b} \rangle & = & 
                \pm \frac{8}{3} r_2 p^0 p^1               \quad,\nonumber \\
 \langle \hat{O}^{(M)}_{r3}   \rangle & = & 
             \pm \frac{2}{3} r_3 p^1 \left( (p^1)^2 + 2 (p^0)^2 \right)
             \quad.
  \label{operes}
\end{eqnarray}

\section{Smearing}
\label{appSmearing}

\def\gsim{\mathrel{\rlap{\lower4pt\hbox{\hskip1pt$\sim$}}
    \raise1pt\hbox{$>$}}}                
The method we use for smearing is to
smear the quark in a plane $x_4=t$ \cite{Alexandrou}:
\begin{equation}
   {}^S\psi_{f\alpha}^{a}(\vec{x},t) =
         \sum_{\vec{y}} {}^SH^{ab}(\vec{x},\vec{y};U,t)
           \psi_{f\alpha}^{b}(\vec{y},t) \,,
\end{equation}
where the kernel $H$ is chosen to have the correct gauge transformation
properties
and is diagonal in spin space. $S$ is the smearing label.
So for example for no smearing $S=L=\mbox{local}$,
then ${}^LH^{ab}(\vec{x},\vec{y};U,t)=\delta^{ab}\delta_{\vec{x}\vec{y}}$.
${}^SH$ is also taken as Hermitian:
\begin{equation}
      {}^SH^{ba}(\vec{y},\vec{x};U,t)^* =
      {}^SH^{ab}(\vec{x},\vec{y};U,t) \,.
\end{equation}
Also the smeared anti-quark is defined as
\begin{equation}
   {}^S\bar{\psi}_{f\alpha}^{a}(\vec{x},t) =
         \sum_{\vec{y}} \bar{\psi}_{f\alpha}^{b}(\vec{y},t)
          H^{ba}_S(\vec{y},\vec{x};U,t) \,.
\end{equation}
Note that we can choose different smearing for the quark and anti-quark.
Thus for a smeared meson operator we have
\begin{equation}
   {}^{S^\prime S}\eta(t,\vec{p}) = \sum_{\vec{x}} F_{ff'}
               e^{-\I\vec{p}\cdot\vec{x}} \, 
    {}^{S^\prime}\!\bar{\psi}_{f\alpha}^{a}(\vec{x},t) \Gamma_{\alpha\beta}
           \, {}^S\psi_{f'\beta}^{a}(\vec{x},t)
\end{equation}
with the appropriate correlation function
\begin{equation}
   C^{S^\prime S}(t,\vec{p};t_0) = \langle {}^{S^\prime S'}\eta(t,\vec{p})
                           \,    {}^{S S}\eta(t_0,-\vec{p})
                                   \rangle
\end{equation}
so that $S$ and $S^{\prime}$ is smearing at the source, sink respectively.

The smeared quark propagator is defined by
\begin{equation}
   {}^{S^\prime S}G_{\alpha\beta}^{fab}(x,y;U)\delta_{ff^\prime} =
        \langle {}^{S^\prime}{\psi}_{f\alpha}^{a}(x)
                {}^{S}\bar{{\psi}}_{f'\beta}^{b}(y) \rangle
                 _{\rm fermions}
\end{equation}
(${}^{LL}G \equiv G$). So in meson correlation functions we can
simply replace $G$ with ${}^{{S^\prime}S}G$ to allow for smearing.
 
The smeared quark propagators are found sequentially:
\begin{itemize}
   \item Generate the smeared source ${}^SS$ from a point source
         at $(\vec{x}_0,t_0)$ and so with
         $S_{0\alpha}^a(\vec{x}t) = \delta_{\vec{x}\vec{x}_0}\delta_{tt_0}
                           \delta^{aa_0}\delta_{\alpha\alpha_0}$,
         \begin{equation}
            {}^SS_{\alpha \alpha_0}^{faa_0}(\vec{x},t;\vec{x}_0,t_0) =
            {}^SH^{aa_0}(\vec{x},\vec{x}_0;U,t)\delta_{tt_0}
                                               \delta_{\alpha\alpha_0}\,.
         \label{smear1}
         \end{equation}
   \item Find ${}^{LS}G$, by solving $M{}^{LS}G = {}^{S}S$.
         We thus have
         \begin{equation}
            {}^{LS}G_{\alpha\beta}^{fab}(\vec{x},t;\vec{x}_0,t_0)
                   \equiv \sum_{\vec{y}} 
                 G_{\alpha\beta}^{fab^\prime}(\vec{x},t;\vec{y},t_0)
                 {}^SH^{b^\prime b}(\vec{y},\vec{x}_0;U,t_0) \,.
         \end{equation}
   \item From ${}^{LS}G$ we generate ${}^{S^\prime S}G$ by applying
         ${}^{S^\prime}H$:
         \begin{equation}
          {}^{S^\prime S}G_{\alpha\beta}^{fab}(\vec{x},t;\vec{x}_0,t_0) = 
          \sum_{\vec{y}}
          {}^{S^\prime}H^{aa^\prime}(\vec{x},\vec{y};U,t)
          {}^{LS}G_{\alpha\beta}^{fa^\prime b}
                 (\vec{y},t;\vec{x}_0,t_0) \,.
         \end{equation}
         Note that this step can be expensive (in CPU time) 
         in comparison to eq.~(\ref{smear1}) as we must smear on every
         $x_4=t$ plane.
\end{itemize}

Practically we shall use Jacobi smearing (as advocated mainly by
\cite{UKQCD}). This is given by
\begin{equation}
 \sum_{\vec{x}^\prime} K(\vec{x},t;\vec{x}^\prime t)\, 
    {}^SS(\vec{x}^\prime ,t) = S_0(\vec{x},t) \,,
\label{ksmear}
\end{equation}
where $S_0$ is the original point source. Here
\begin{equation}
   K = 1 - \kappa_s D_s
\end{equation}
and 
$D_s$ is a covariant derivative
in the $x_4=t$ plane, viz.
\begin{equation}
   D_{s\alpha\beta}^{ab}(\vec{x},t;\vec{y},t) = \delta_{\alpha\beta} 
        \sum_{i=1}^{3}
         \left[ U_i^{ab}(\vec{x},t)\delta_{\vec{x}+\vec{\imath},\vec{y}} + 
                U_i^{\dagger ab}(\vec{x}-\vec{\imath},t)
                      \delta_{\vec{x}-\vec{\imath},\vec{y}} \right] \,.
\end{equation}
Hence we need $H = K^{-1}$.
Rather than performing this inversion
completely we Jacobi iterate $N_s$ times, so
\begin{equation}
   {}^SS^{(n)}(\vec{x},t) = 
        S_0(\vec{x},t) + \kappa_s D_s {}^SS^{(n-1)}(\vec{x},t)
                     \qquad n = 1, 2, \ldots
\end{equation}
with ${}^SS^{(0)}(\vec{x},t) = S_0(\vec{x},t)$. 

We thus have two parameters, $\kappa_s$, $N_s$ at our disposal. $\kappa_s$
controls the coarseness of the iteration, while increasing $N_s$
increases the size of the smeared object roughly like a random walk.
Physically we wish to smear until our source is about the size of the meson.
A suitable measure of the (rms) radius is given by
\begin{equation}
   r^2 = {\sum_{\vec{x}} \left( \vec{x} - \vec{x}_0 \right)^2
           \left| {}^SS (\vec{x},t_0; \vec{x}_0,t_0) \right|^2
         \over
           \left| {}^SS (\vec{x},t_0; \vec{x}_0,t_0) \right|^2 } \,.
\end{equation}
(Note that on a periodic lattice, $(\vec{x}-\vec{x}_0)^2$ is taken as the
minimum distance from $\vec{x}$ to $\vec{x}_0$.)
Explicitly for $\beta =6.0$ we have chosen $\kappa_s=0.21$, $N_s=50$.
This gives $ra$ of about $3.5a \gsim 0.5\mbox{fm} $ 
which
corresponds roughly to the hadron radius.

\end{document}